\documentclass[a4paper,12pt]{article}
\usepackage{jcappub}
\usepackage[utf8]{inputenc}
\usepackage[T1]{fontenc}
\usepackage[spanish,english]{babel}
\usepackage{graphicx}
\usepackage{amsmath}
\usepackage{amssymb}
\usepackage{url}
\usepackage{verbatim}
\usepackage{hyperref}
\usepackage[dvipsnames]{xcolor}
\usepackage{etoolbox}

\newcommand{\dd}{\text{d}}

\definecolor{changegreen}{RGB}{0,160,80}
\definecolor{changered}{RGB}{180,0,0}

\title{\boldmath The impact of gamma-ray propagation effects on indirect dark matter searches}
\author[a, b]{Ignacio Martínez López,}
\author[a, c, d,e]{Rafael {Alves~Batista},}
\author[a, c]{Miguel~A. {Sánchez-Conde},}
\author[b]{Antonio Juan Rubio-Montero}
\affiliation[a]{Departamento de Física Teórica, Mod. 15, Universidad Autónoma de Madrid,\\ E-28049 Madrid, Spain.}
\affiliation[b]{Dept. de Tecnología, Centro de Investigaciones Energéticas, Medioambientales y Tecnológicas (CIEMAT), Av. Complutense 40, Madrid, Spain}
\affiliation[c]{Instituto de Física Teórica, IFT UAM-CSIC,\\ Calle Nicolás Cabrera 13-15, Campus de Cantoblanco, E-28049 Madrid, Spain.}
\affiliation[d]{Institut d'Astrophysique de Paris, Sorbonne Université, CNRS UMR 7095, 98 bis bd Arago, 75014 Paris, France}
\affiliation[e]{Laboratoire de Physique Nucléaire et des Hautes Énergies, Sorbonne Université, 4 place Jussieu, 75005 Paris, France}

\emailAdd{ignacio.martinez@ciemat.es}
\emailAdd{rafael.alves\_batista@iap.fr}
\emailAdd{miguel.sanchezconde@uam.es}
\emailAdd{antonio.rubio@ciemat.es}

\abstract{In this work, we investigate dark matter (DM) detection in the context of weakly interacting massive particles (WIMPs). Upon annihilation, WIMPs generate cascades of secondary particles through various channels, many of which culminate in the production of gamma rays. As these gamma rays travel toward Earth, their spectra are reshaped by interactions with the intervening medium. While current models typically account for attenuation via pair production on the extragalactic background light, they often neglect the fate of the resulting electrons and positrons, specifically subsequent inverse Compton scattering of these secondary particles, which can regenerate high-energy gamma rays.
Here, we revisit the predicted gamma-ray fluxes from WIMP annihilation by performing a more detailed treatment of propagation effects. We show that for distant sources and annihilation channels such as $\tau^+\tau^-$, the full treatments can significantly alter the observed gamma-ray flux, by up to three orders of magnitude for heavy WIMPs. This has an impact on current dark matter limits derived without taking into account propagation effects, depending on the considered WIMP mass and annihilation channel. Our study demonstrates the importance of a detailed propagation treatment for indirect dark matter searches, and the need to account for such effects in order to obtain  accurate, more reliable dark matter signal predictions and exclusion limits.}
\arxivnumber{2603.20409}

\begin{document}
\maketitle
\flushbottom

\section{Introduction}
\label{sec: Introduction}

Unveiling the nature of dark matter (DM) remains one of the central challenges in modern cosmology and particle physics. While a wide array of astrophysical observations -- from galactic rotation curves to gravitational lensing and cosmic microwave background anisotropies -- confirm its presence, the particle identity of DM is still unknown~\cite{Bertone_2005, Bertone_2018}. 

Among the most extensively studied candidates are weakly interacting massive particles (WIMPs), motivated, among other reasons, by the so-called ``WIMP miracle'': the observation that particles with masses and interaction cross sections typical of the electroweak scale naturally yield a relic abundance consistent with current measurements~\cite{GONDOLO1991145,Griest91,SREDNICKI1988693,STEIGMAN1985375}. Although WIMPs do not arise within the Standard Model, they naturally appear in a wide range of theories beyond it and remain one of the benchmark scenarios for indirect dark matter searches~\cite{Bertone_2005,Arcadi_2018,Roszkowski_2018}. A historically important example is supersymmetry, where the lightest neutralino provides a prototypical WIMP candidate~\cite{HABER198575}. However, the propagation effects investigated in this work do not depend on any specific particle-physics realization and are expected to apply to any dark matter scenario producing comparable gamma-ray and electron-positron spectra.

Given the lack of conclusive evidence from collider~\cite{boveia2018a, lee2019a} and direct detection experiments~\cite{misiaszek2024a} so far, indirect detection has become an increasingly important avenue to test the WIMP paradigm. This approach seeks astrophysical signals resulting from DM annihilation or decay in regions of high density~\citep{Gaskins_2016,Leane:2020liq}, such as galactic centres~\cite{Ajello_2016,Ackermann_2017,DiMauro2021mar,Zuriaga-Puig_2023}, dwarf galaxies~\cite{Ackerman2015nov,armand2021,Gammaldi2021,mcdaniel2023,fernandezsuarez2025}, galaxy clusters~\cite{Ackermann_2015oct,DiMauro2023}, or dark satellites~\cite{Ackermann_2012,Coronado-Blazquez2019jul,Coronado-Blazquez2019nov,Coronado-Blazquez2022}, where the expected DM signal can be enhanced. Observations are carried out using space-based telescopes like the Fermi Large Area Telescope (Fermi-LAT)~\cite{Atwood_2009, Conrad_2015, Charles_2016}, as well as ground-based Imaging Atmospheric Cherenkov Telescopes (IACTs)~\cite{WEEKES2002221,LORENZ2004339,HINTON2004331,Abeysekara_2017,Acharya2018,Ma_2022,Mukherjee_2024,Abreu_2025swgo}, providing complementary coverage over a broad energy range.

Regarding the astrophysical messengers used for indirect DM detection, gamma rays are particularly useful because, unlike charged cosmic rays, they are not deflected by magnetic fields and thus retain directional information about their sources~\cite{Morselli2023,fernandezsuarez2025}. Cosmic rays provide complementary constraints on high-energy astrophysical processes~\cite{Salati2023}. Because neutrinos can escape from dense regions due to their weak interactions, they offer a unique probe of obscure environments potentially associated with DM~\cite{Abe_2020,Albert_2020,Aiello_2025}, despite the difficulty in their detection.

The complementarity of current gamma-ray instruments allows for a broad coverage of the energy range relevant to DM searches. IACTs, such as MAGIC, VERITAS and H.E.S.S., provide leading constraints at multi-TeV energies \citep{Archambault_2017,Abdallah_2018}. In particular, the MAGIC collaboration has derived upper limits on the annihilation cross section into monochromatic photons at the level of
$\langle \sigma v \rangle \lesssim 5 \times 10^{-28} \; \mathrm{cm}^3 \, \mathrm{s}^{-1}$
at 1~TeV and
$\langle \sigma v \rangle \lesssim 1 \times 10^{-25} \; \mathrm{cm}^3 \, \mathrm{s}^{-1}$
at 100~TeV~\citep{Abe_2023}.

At lower masses, the most robust and constraining limits are currently provided by Fermi-LAT observations of dwarf spheroidal galaxies, which reach the thermal relic cross section, 
$\langle \sigma v \rangle \sim 3 \times 10^{-26} \; \text{cm}^3 \, \text{s}^{-1}$ 
around $m_{\chi} \sim 100~\text{GeV}$ for several annihilation channels~\citep{mcdaniel2023}. For heavy WIMPs, ground-based wide field-of-view observatories such as LHAASO and HAWC further extend sensitivity to DM candidates in the multi-TeV to PeV range~\citep{Albert_2018,Cao_2022}.

Looking ahead, the forthcoming Cherenkov Telescope Array Observatory (CTAO) is expected to significantly advance these efforts~\citep{Acharya2018}. Designed as a next-generation ground-based observatory with sites in both hemispheres, CTAO will greatly improve upon the capabilities of current instruments by offering enhanced sensitivity, wider energy coverage (from a few tens of GeV up to several hundreds of TeV), improved angular and energy resolution, and faster response times. These advancements will enable more stringent tests of DM models and provide broader opportunities for discovery in the very-high-energy regime.

While ground and space-based instruments provide critical observational constraints on DM through the detection of high-energy gamma rays, interpreting these observations requires a detailed understanding of how such photons propagate through the Universe. Many DM annihilation channels produce gamma rays that must traverse the intergalactic medium before reaching our detectors. Along the way, these high-energy photons interact with pervasive radiation fields 
(for a recent comprehensive review see~\citep{driver2021}), 
including the cosmic microwave background (CMB)~\citep{mather1994}, 
extragalactic background light (EBL)~\citep{franceschini2008}, 
and cosmic radio background (CRB)~\citep{fixsen2011}. These interactions reshape the energy spectrum originally emitted by the DM source, particularly at the highest energies. Therefore, a robust interpretation of observational data hinges on accurately modelling the effects of these backgrounds on gamma-ray propagation. 

Previous studies have already highlighted the importance of electromagnetic cascades in dark matter searches. In particular, cascade reprocessing has been shown to significantly strengthen constraints on very heavy dark matter, especially in decaying scenarios and for cluster environments \cite{MuraseBeacom2012,Song:2023xdk}. These works demonstrate the relevance of secondary gamma-ray production, but a detailed characterization of the propagated gamma-ray spectrum itself remains essential for robust predictions in annihilating dark matter scenarios.

The main goal of the present work is to address the impact of gamma-ray propagation effects on indirect DM searches. Indeed, most indirect DM searches consider only gamma-ray attenuation by EBL photons, neglecting secondary gamma-ray production via inverse Compton scattering. In this work, we investigate the impact of including both pair production and inverse Compton scattering with CMB and EBL, and show how this modifies the predictions for DM annihilation signals in the gamma-ray energy range.

We structure our analysis as follows. In Section~\ref{sec: gamma-ray signals}, we describe the methods used to characterise the intrinsic gamma-ray spectrum generated by DM annihilation and outline the computational approach employed using CosmiXs~\citep{Arina2024}. Section~\ref{sec: Propagation} then presents the propagation framework based on CRPropa~3~\citep{Batista_2016, Alves_Batista_2022}, which incorporates the main physical processes affecting gamma rays in transit. Our results are presented in Section~\ref{sec: Results}, including a general scenario (Subsection~\ref{subsec: Generic}), a specific case involving the Perseus galaxy cluster (Subsection~\ref{subsec: Perseus}), and the resulting impact on constraints set on DM particle properties (Subsection~\ref{subsec: DM limits}). Finally, we discuss the broader implications of these findings in Section~\ref{sec: Discussion}.

\section{Dark matter-induced gamma-ray signals}
\label{sec: gamma-ray signals}

In this section, we characterise the flux of gamma rays resulting from DM annihilation, which primarily occurs in DM-dense regions such as galactic haloes and galaxy clusters. Note that we focus solely on WIMP annihilation and ignore the flux from WIMP decay because we expect the cross section of WIMPs to correspond to the thermal freeze-out \cite{Dodelson:2003ft,Steigman_2012}, while in the case of decay we have no clues of the corresponding interaction rate (although approaches are made to the decay lifetime \cite{Arina_2010,Acciari_2018}); thus, annihilation is currently considered a more viable approach to probe the nature of DM.

The gamma-ray flux from DM annihilation, observed within a solid angle $\Delta \Omega$, is given by~\cite{Bergstr_m_1998,Coronado-Blazquez2019jul}:
\begin{equation}
\label{eq: ann and prop}
F = \left(\frac{x\langle \sigma v\rangle}{8 \pi} \frac{1}{m^2_{\chi}} \sum_f \mathcal{B}_f \int\limits_{E_\text{th}}^{E}\frac{\dd N^\text{f}_{\gamma}}{\dd E_{\gamma}}\, \dd E_{\gamma} \right)J_T \,,
\end{equation}
where $x = 1$ if DM is a Majorana particle and $x = 1/2$ otherwise. Here, $\langle \sigma v\rangle$ denotes the velocity-averaged annihilation cross section for DM particles with mass $m_{\chi}$, $\frac{\dd N^\text{f}_{\gamma}}{\dd E_{\gamma}}$ is the gamma-ray yield per annihilation to a channel $\text{f}$, and $\mathcal{B}_\text{f}$ is the corresponding branching ratio.

The annihilation process is schematically represented as:
\begin{equation}
\label{eq: channel}
    \chi \chi \rightarrow \underbrace{\left[X_1 X_2 ... X_N\right]}_{\text{intermediate states}} \rightarrow \overbrace{\left(Y_{11} ... Y_{1{a_1}}\right)... \left(Y_{N1} ... Y_{N{a_N}}\right)}^{\text{stable particles}},
\end{equation}

where $\chi$ represents DM particles, the intermediate states denote annihilation channels, and the resulting stable products correspond to the particles that give rise to the signals targeted in indirect DM searches. We assume that WIMPs predominantly annihilate through a single final state, such that $\mathcal{B}_\text{f} = 1$. The limits of integration correspond to the detector's energy threshold ($E_{\text{th}}$), and to an upper limit E, selected based on the highest DM mass under consideration. According to recent studies~\cite{Frumkin_2023}, viable WIMP candidates could reach masses up to 150~TeV, consistent with the thermal relic abundance framework. This sets the relevant energy scale for the upper limit of the integral in Eq.~\ref{eq: ann and prop}, since the energy of the resulting gamma rays depends on the mass of the DM particle.

Although these terms encode the microscopic properties of DM interactions, the total flux also depends on the astrophysical distribution of DM along the line of sight. The second multiplicative term in Eq.~\ref{eq: ann and prop}, the so-called astrophysical $J$-factor, captures this dependence, as it reflects the square of the DM density ($\rho_\text{DM}(r)$) integrated over the source volume and normalised by its distance squared, $D^2$:

\begin{equation}
\label{eq: J}
J_T = \dfrac{1}{D^2} \int\limits_{V} \rho_\text{DM}^2(r) \, \dd V.
\end{equation}

The DM density profile must satisfy constraints based on the nature of DM to realistically describe its distribution in halos. Since the 1990s, the Navarro-Frenk-White (NFW) profile~\citep{Navarro_1996} has been widely adopted, though improved N-body cosmological simulations have suggested alternative profiles, depending on the precise halo (or subhalo) characteristics, e.g.~\citep{Einasto_1965,Hernquist_1990,Retana_Montenegro_2012,DiCintio_2013,Errani_2021}.

\subsection{CosmiXs: gamma-ray flux from DM annihilation}
\label{subsec: CosmiXs}

We employ the CosmiXs code~\cite{Arina2024} to compute the gamma-ray spectra resulting from WIMP annihilation. CosmiXs builds upon prior work~\citep{Marco_Cirelli_2011, Ciafaloni_2011, Fischer_2016, Backovi__2014} using quantum field theory (QFT) and Monte Carlo algorithms~\citep{Amoroso_2019}. It was designed to calculate the energy spectra of secondary particles resulting from DM annihilation or decay, incorporating quantum electrodynamics (QED) and quantum chromodynamics (QCD) effects.

We use the publicly available tables within CosmiXs~\citep{Arina2024} to obtain the spectra from DM annihilation for WIMP masses ranging from 5~GeV to 100~TeV. This covers the parameter space for cold DM consistent with thermal freeze-out in the early universe. More precisely, the prompt photon spectrum, $\frac{\mathrm{d}N_{\gamma}^{f}}{\mathrm{d}E_{\gamma}}$, entering the factor in parentheses in Eq.~\ref{eq: ann and prop} is extracted from the tables provided by CosmiXs. Based on CosmiXs simulations, this spectrum serves as the spectral component of gamma-ray signal templates for observatories such as the previously mentioned Fermi-LAT and IACTs; the spatial component is determined by the astrophysical DM distribution.

\begin{figure}[ht!]
\centering
\includegraphics[width=0.495\textwidth]{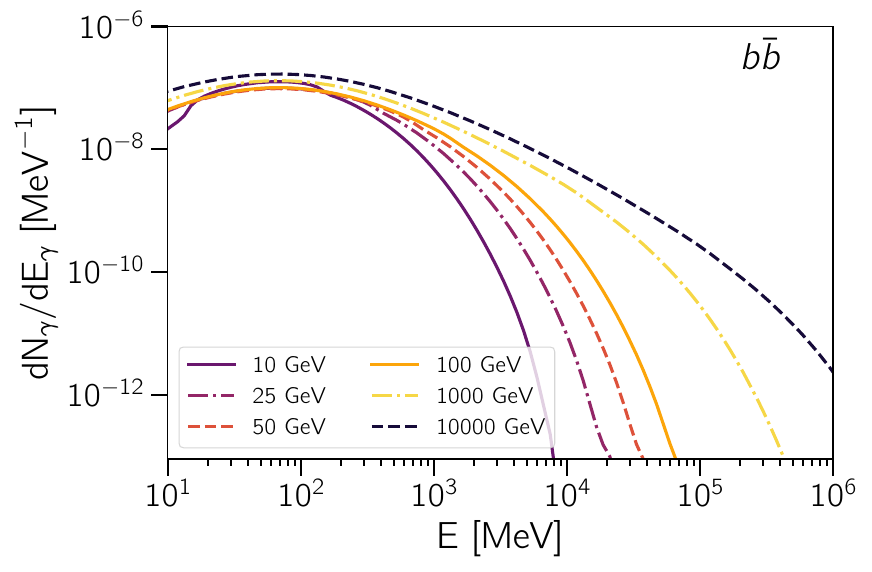}
\includegraphics[width=0.495\textwidth]{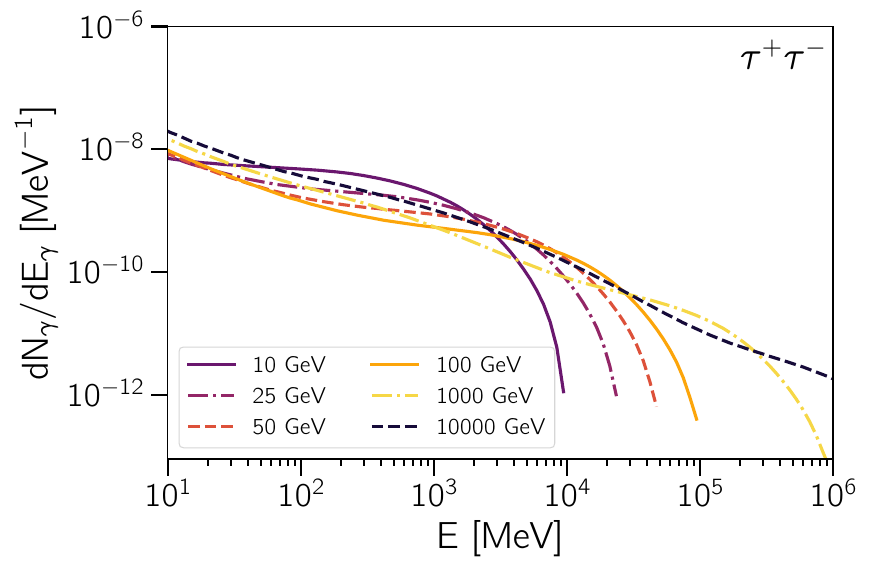}
\caption{Spectrum of gamma rays produced through the $b\bar{b}$ (left) and $\tau^{+}\tau^{-}$ (right panel) annihilation channels ($dN_\gamma / dE_\gamma$), as a function of the energy, for various DM masses.}
\label{fig: CosmiXs}
\end{figure}

Our analysis focuses on the $b\bar{b}$ and $\tau^{+}\tau^{-}$ channels, which are commonly studied alongside the $W$ channel. Figure~\ref{fig: CosmiXs} displays the resulting spectra for multiple WIMP masses within these channels. CosmiXs is used exclusively to compute the prompt particle spectra at the source. The other quantities entering Eq.~\ref{eq: ann and prop}, including the astrophysical $J$-factor and the annihilation cross section, are adopted from dedicated studies in the literature and implemented consistently in our analysis.

\section[Propagation effects on gamma rays from DM annihilation]{Impact of propagation effects on gamma-ray detection from DM annihilation}
\label{sec: Propagation}

The annihilation of DM particles in a hypothetical dense concentration of WIMPs in the Universe can produce a wide variety of particles. Among them, cosmic rays, neutrinos, and gamma rays stand out. In this work, we focus on gamma-ray detection due to the higher sensitivity of our instruments to them and the fact that gamma rays, being electrically neutral, can be traced back to their sources. 

However, detecting gamma rays alone is not enough. Interpreting the observed flux requires understanding how these photons evolve on their journey through the Universe. Thus, to fully characterise the energy spectrum of gamma rays from DM annihilation, it is essential to account for propagation effects. As gamma rays travel from the source to Earth, they may interact with pervasive background radiation fields, resulting in a lower flux observed by detectors. Moreover, the adiabatic expansion of the Universe also incurs energy losses.

One of the primary attenuation mechanisms is pair production~\cite{breit1934a} via interactions with background photons, notably from the CMB, EBL and CRB:
\begin{align}
\label{eq: pair prod}   
\gamma + \gamma_\text{bg} \rightarrow e^+ + e^- .
\end{align}
These produced charged leptons can subsequently undergo inverse Compton scattering~\cite{klein1929}, transferring energy back to background photons:
\begin{align}
\label{eq: subsequent}   
e^\pm + \gamma_\text{bg} \rightarrow e^\pm + \gamma.
\end{align}
The successive occurrence of these interactions forms an electromagnetic cascade.

While pair production is often regarded as a pure attenuation process, the resulting electrons and positrons retain a significant fraction of the energy of the primary gamma ray. These secondary leptons subsequently upscatter background photons through inverse Compton scattering, generating new gamma rays that may undergo further interactions. The resulting electromagnetic cascade redistributes energy from the highest energies towards lower-energy photons, partially regenerating the observable gamma-ray signal.

Additionally, higher-order interactions such as double pair production~\cite{Cheng1970}:
\begin{align}
\label{eq: double pair prod}   
\gamma + \gamma_\text{bg} \rightarrow e^+ + e^- + e^+ + e^-.
\end{align}
and triplet pair production~\cite{Bonometto1972}:
\begin{align}
\label{eq: triplet pair prod}   
e^{\pm} + \gamma_\text{bg} \rightarrow e^{\pm} + e^+ + e^-.
\end{align}
can occur, but with a negligible effect in the energy range we study in this work~\cite{Heiter_2018}.

The likelihood of such interactions is strongly energy-dependent and is therefore tied to the DM particle mass. This dependence is quantified through the inverse of the mean free path ($\lambda^{-1}$), which represents the interaction probability per unit distance for the incident particle with the background fields~\citep{Alves_Batista_2021}. 

\begin{figure}[htb]
\centering
\includegraphics[width=0.6\textwidth]{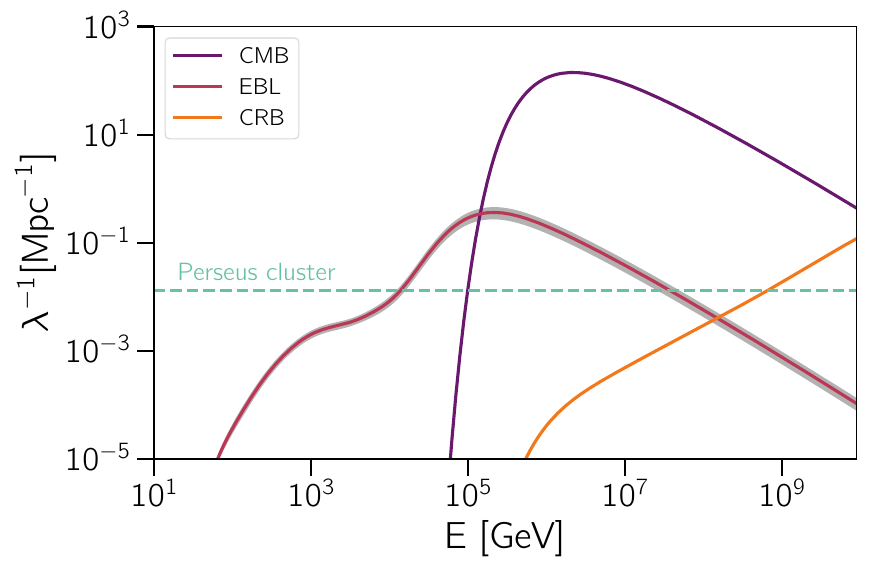}
\caption{Inverse mean free path ($\lambda^{-1}$) of gamma rays as a function of energy. The purple curve represents interactions with the CMB. The red shaded band represents the uncertainty range of the EBL model~\citep{Saldana_Lopez_2021}, corresponding to the lower and upper limits derived from observational constraints. The central line indicates the best-fit model, and the contribution from the CRB is also shown in orange. Data from ref~\citep{NITU2021102532}. The blue dashed line indicates the inverse distance corresponding to the redshift of the Perseus Cluster ($z \sim 0.017$)~\citep{kang2024_perseus}. This distance is highlighted due to its role in later sections, where it is used to constrain WIMP properties.}
\label{fig: free path PP}
\end{figure}

Figure~\ref{fig: free path PP} shows $\lambda^{-1}$ for the pair-production process in Eq.~\ref{eq: pair prod} as a function of gamma-ray energy for interactions with CMB, EBL, and CRB, along with the reference distance to the Perseus cluster. This specific distance is highlighted because it corresponds to the source from which we later derive constraints on the WIMP mass and annihilation cross section as an example of the impact of gamma-ray propagation effect for DM searches.
The CRB contribution is shown for completeness but becomes relevant only at ultra-high energies, well above the range considered here, and is therefore neglected in our simulations. The CMB is known with extremely high precision and introduces negligible uncertainties. The dominant uncertainty arises from the EBL, for which observationally motivated uncertainty bands have been derived using modern reconstruction methods based on galaxy surveys and gamma-ray observations~\citep{Saldana_Lopez_2021}. As shown by the red shaded region in Figure~\ref{fig: free path PP}, this uncertainty remains small over the energy range relevant to this work and is therefore expected to have only a minor impact on the electromagnetic cascade.

\begin{figure}[htb]
\centering
\includegraphics[width=0.6\textwidth]{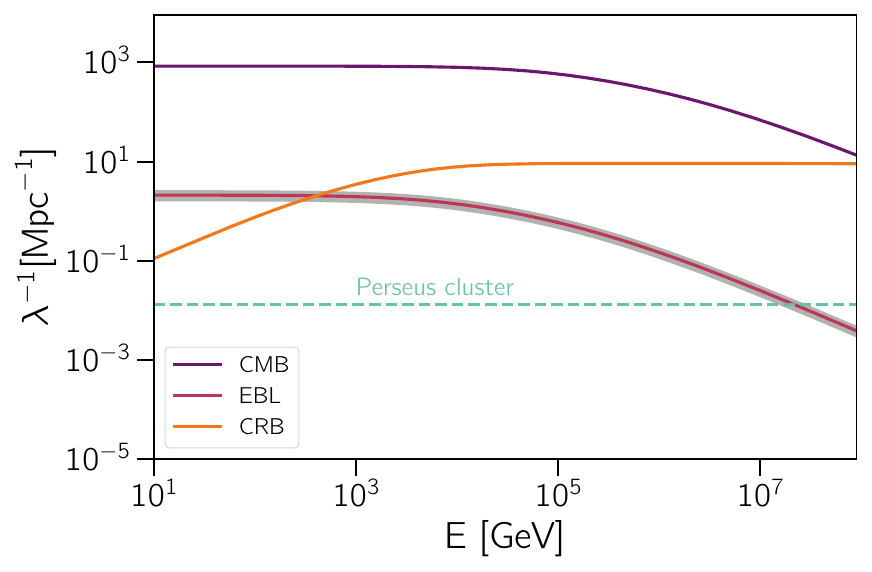}
\caption{Same as Figure~\ref{fig: free path PP}, but for inverse Compton scattering between high-energy electrons and background photons from the CMB, EBL, and CRB. The inverse distance to the Perseus cluster is included for reference, as it serves as a benchmark in the analysis of WIMP-induced gamma-ray fluxes presented in later sections.}
\label{fig: free path ICS}
\end{figure}

Figure~\ref{fig: free path ICS} shows the inverse mean free path for inverse Compton scattering, described by Eq.~\ref{eq: subsequent}, as a function of electron energy. As in the case of pair production, the uncertainty band remains relatively small across the energy range relevant to this work and is therefore not expected to significantly affect the development of the electromagnetic cascade.
The inverse mean free path for both pair production and inverse Compton scattering, illustrated in Figures~\ref{fig: free path PP} and~\ref{fig: free path ICS}, respectively, can be written as~\citep{Alves_Batista_2021}:
\begin{align}
\label{eq: attenuation}
\lambda^{-1}(E,z) = \frac{1}{8 \beta E^2}\int\limits_0^{\infty}\int\limits_{s_\text{min}}^{s_\text{max}} \frac{1}{\epsilon^2}\frac{\dd n(\epsilon,z)}{\dd \epsilon}  (s - m^2 c^4) \sigma\left(s\right) \, \dd s  \dd\epsilon \,,
\end{align}

where $\epsilon$ is the energy of the background photon, $\sigma$ is the cross section of the interaction, and $m$ is the mass of the energetic particle that interacts with these photons. In the case of pair production, $s_\text{min} = 4m_e^2c^4$ and $s_\text{max} = 4E\epsilon$. For inverse Compton scattering, $s_\text{min} = m_e^2c^4$ and $s_\text{max} = m_e^2c^4 + 2E\epsilon(1+ \beta)$, with $\beta$ the speed of the electron in units of $c$.

Most indirect DM searches consider only attenuation from pair production with EBL photons, neglecting secondary gamma-ray production via inverse Compton scattering. In this work, we investigate the impact of including pair production and inverse Compton scattering with CMB and EBL and show how it modifies our predictions for gamma-ray fluxes from DM annihilation.

\subsection{CRPropa: treatment of propagation effects}
\label{subsec: CRPropa}

To model the propagation of gamma rays and their interactions with the cosmological environment, we use the CRPropa code~\citep{Batista_2016, Alves_Batista_2022}. CRPropa is a Monte Carlo simulation framework designed to track high-energy astrophysical messengers, including cosmic rays, neutrinos, electrons, and gamma rays, as they propagate through intergalactic space. 
CRPropa supports both one-dimensional (1D) and three-dimensional (3D) propagation. In this work, we use the 1D mode to model interactions and energy-loss processes while neglecting magnetic field effects~\cite{Batista_2016}. This approximation is appropriate for the highly relativistic particles considered here, for which inverse Compton photons are emitted nearly along the direction of the parent electrons, while magnetic-field effects are not expected to significantly alter the resulting gamma-ray spectra \cite{Blumenthal1970, Alves_Batista_2021}.
Following the discussion in Section \ref{sec: Propagation}, pair production is the only gamma-ray attenuation process included in the simulations. Higher-order attenuation mechanisms, such as double and triplet pair production, are neglected because their contribution is expected to be negligible in the energy range considered here~\cite{Heiter_2018}. Secondary electrons and positrons are subsequently propagated including inverse Compton scattering, which together with pair production governs the development of the electromagnetic cascade (see e.g.~\cite{Lee_1998,Alves_Batista_2021}).

To study the full propagation history, we use gamma-ray spectra generated by CosmiXs to reweight the propagation simulations of CRPropa. Each simulation tracks a ``candidate'' particle that evolves through successive steps based on physical modules that simulate interactions, energy losses, and redshift effects. Simulations are terminated when the particle reaches the observer or falls below a minimum energy threshold, set to 1~GeV in our work. 

We simulated approximately $10^6$ gamma-ray events per configuration, with initial energies extending up to 120~TeV, consistent with the upper range of DM particle mass predictions in supersymmetric models. The propagation distances range from 5~Mpc to 600~Mpc. Step sizes vary dynamically between 0.01~kpc and 100~kpc depending on the local interaction rate.
The EBL model used is that of Saldaña-López et al.~\citep{Saldana_Lopez_2021}.

This approach allows us to simulate electromagnetic cascades resulting from DM annihilation, and to evaluate the impact of secondary gamma rays produced via inverse Compton scattering, a contribution often omitted in conventional analyses but shown here to play a crucial role in shaping the observed spectra.

\section{Results}
\label{sec: Results}
This section shows the key findings of our analysis, focusing on the gamma-ray spectra obtained after simulating WIMP-induced signals. We build upon earlier sections by applying the combined framework introduced in this work, linking the gamma-ray yields from WIMP annihilation to their subsequent propagation through cosmological distances. Specifically, we used CosmiXs\footnote{Git commit \href{https://github.com/ajueid/CosmiXs/tree/150786cee5c1bfde77df8711a3c5ac1b362445d8}{150786c}} to generate the prompt spectra for various annihilation channels, and CRPropa 3.2.1\footnote{Git commit  \href{https://github.com/CRPropa/CRPropa3/tree/80bef3b52a89b8000287fd039c669a8d482cd0c6}{80bef3b}} to model the effects of interactions during transit. The resulting spectra, shaped by interactions and energy losses, are presented and discussed with an emphasis on potential observational signatures.

Subsection~\ref{subsec: Generic} focuses on a generic case where the flux is shown in dimensionless units. Subsequently, Subsection~\ref{subsec: Perseus} examines a well-known astrophysical target, the Perseus galaxy cluster. Finally, in Subsection~\ref{subsec: DM limits}, we explore how propagation effects can influence constraints on the expected annihilation cross section of WIMPs, using the Perseus gamma-ray spectra as our example case.

\subsection{Generic study}
\label{subsec: Generic}

To evaluate the role of inverse Compton scattering in the propagation of gamma rays from DM sources, we simulate with CRPropa the gamma-ray flux from two annihilation channels, representative each of them of leptonic and hadronic `families', namely $\tau^{+}\tau^{-}$ and $b\bar{b}$, respectively. We analyse four representative distances: 0.5~Mpc (e.g., NGC 6822, the Barnard’s Galaxy)~\cite{Namumba_2017}, 5~Mpc (e.g., the Sculptor galaxy~\cite{Exp_sito_M_rquez_2022}), 100~Mpc (e.g., the Perseus galaxy cluster~\cite{S_nchez_Conde_2011}), and 600~Mpc (e.g., the Abell 2029 galaxy cluster~\cite{Mirakhor_2021}). As discussed in Subsection \ref{subsec: CRPropa}, we expect the observed spectrum to resemble the initial one for nearby sources. In contrast, gamma rays originating from more distant sources undergo multiple interactions with background photons, such as those from the CMB, EBL, or CRB, which progressively reduce their energy. However, in our simulations, we consider only interactions with the CMB and EBL. At very large distances, many high-energy photons fall below the range of energy we are interested in. This effect becomes more pronounced for higher DM masses, as shown in Figures~\ref{fig: free path PP} and \ref{fig: free path ICS}.

These propagation effects are particularly relevant for DM masses in the TeV range and become negligible in the GeV range unless travelling very long distances. For this reason, we simulate masses $m_\chi = 1, 10, 100$~TeV.

\begin{figure*}[htb!]
\includegraphics[width=0.495\textwidth]{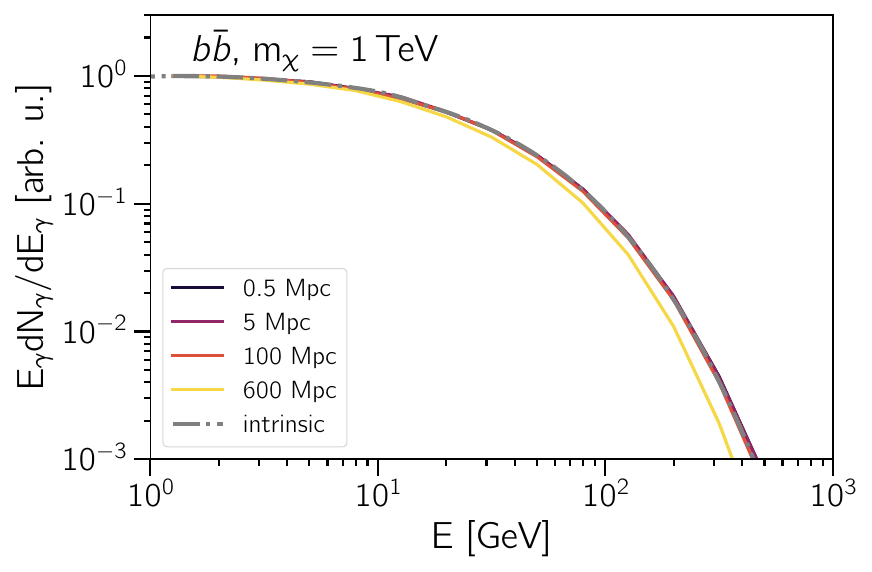}
\includegraphics[width=0.495\textwidth]{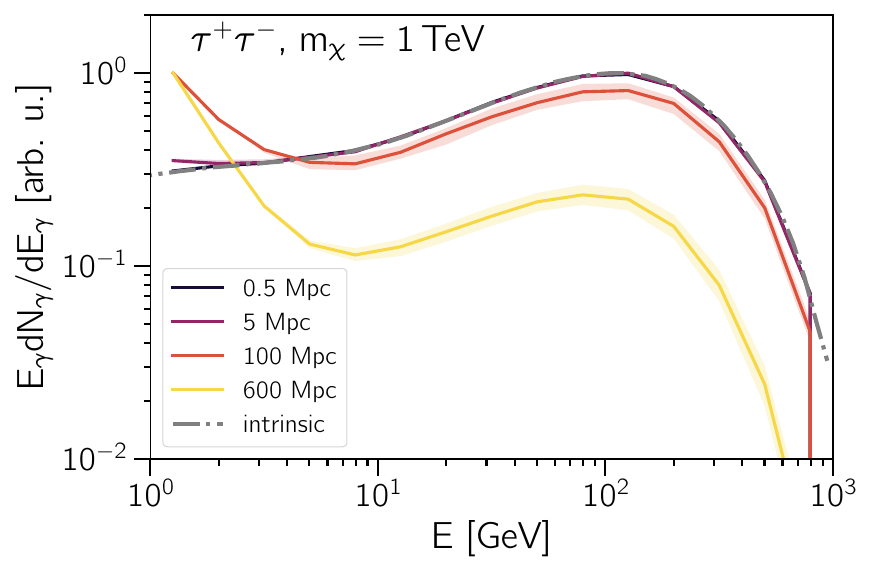}
\includegraphics[width=0.495\textwidth]{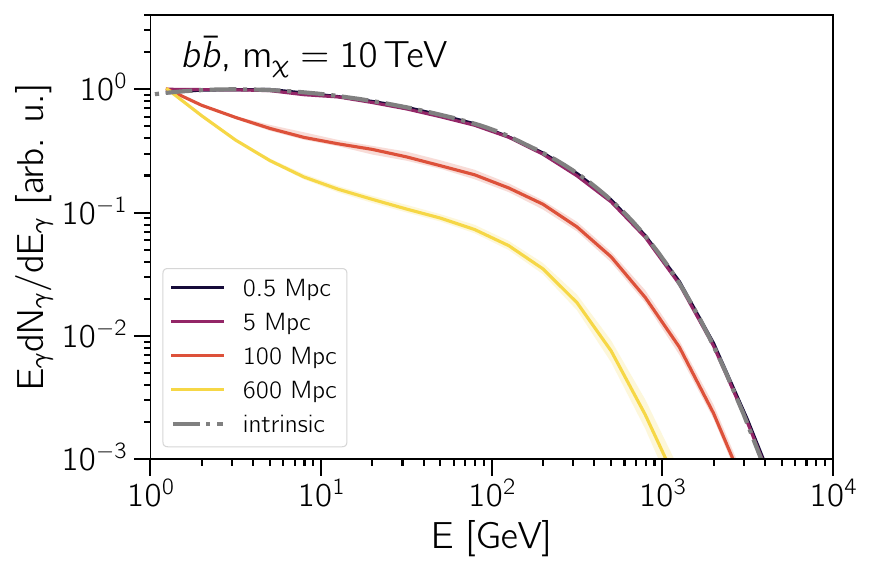}
\includegraphics[width=0.495\textwidth]{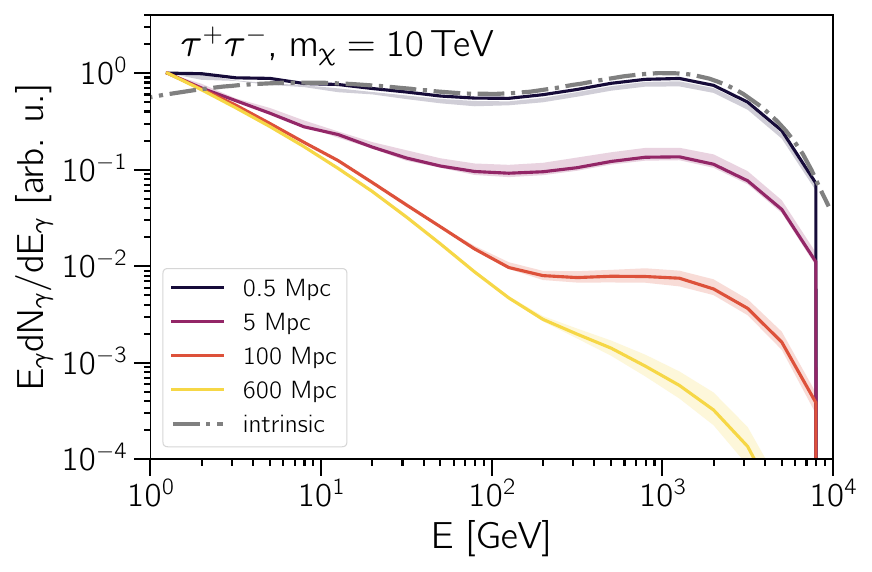}
\includegraphics[width=0.495\textwidth]{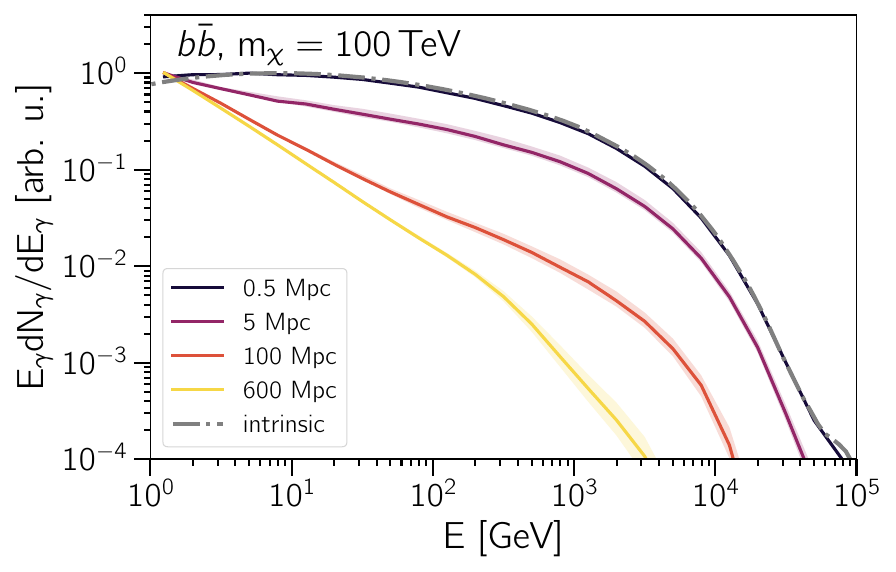}
\includegraphics[width=0.495\textwidth]{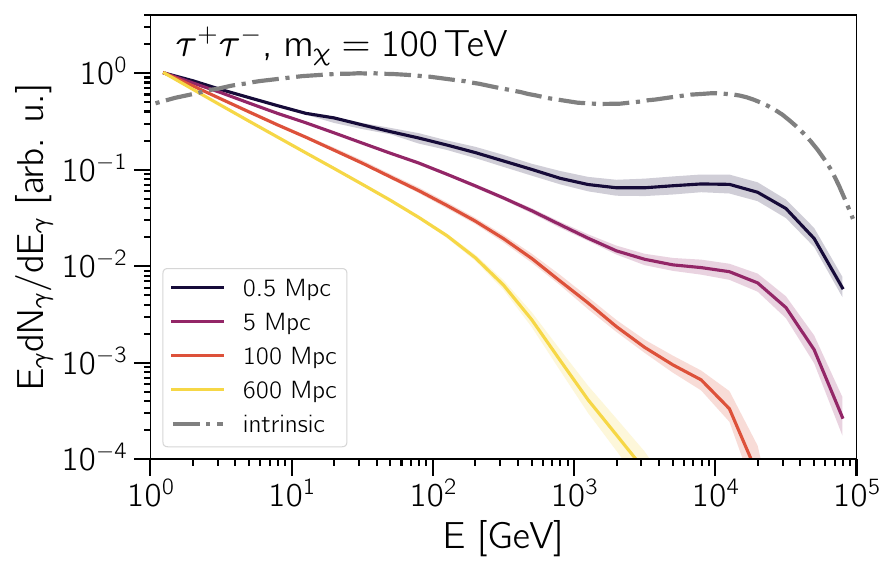}
\caption{Simulated dimensionless gamma-ray flux, expressed as $E\,dN/dE$, as a function of energy (in GeV) for the $b\bar{b}$ annihilation channel (left column) and the $\tau^{+}\tau^{-}$ channel (right column). The panels correspond to DM masses of 1~TeV (upper row), 10~TeV (middle row), and 100~TeV (lower row). The dashed gray lines denote the annihilation spectra at the source position, while the black, purple, red, and yellow solid lines correspond to source distances of 0.5, 5, 100, and 600~Mpc, respectively. The shaded bands indicate the uncertainty associated with the Saldana-Lopez21 EBL model~\cite{Saldana_Lopez_2021} (lower and upper prescriptions), combined with the Monte Carlo statistical uncertainty. All relevant propagation effects are included, in particular inverse Compton scattering, which progressively redistributes the photon energy and modifies the spectral shape with increasing propagation distance.}
\label{fig: Spectrum}
\end{figure*}

\begin{figure*}[htb]
\centering
\includegraphics[width=0.495\textwidth]{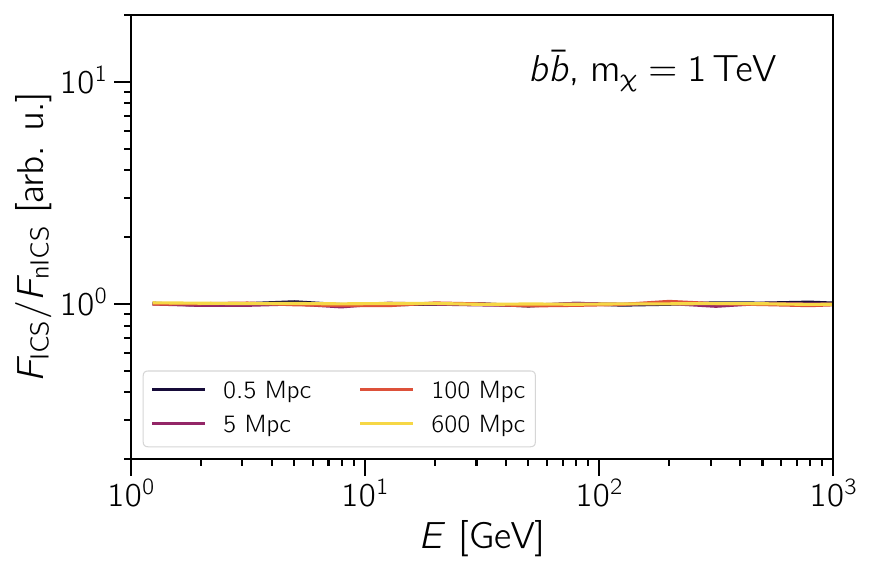}
\includegraphics[width=0.495\textwidth]{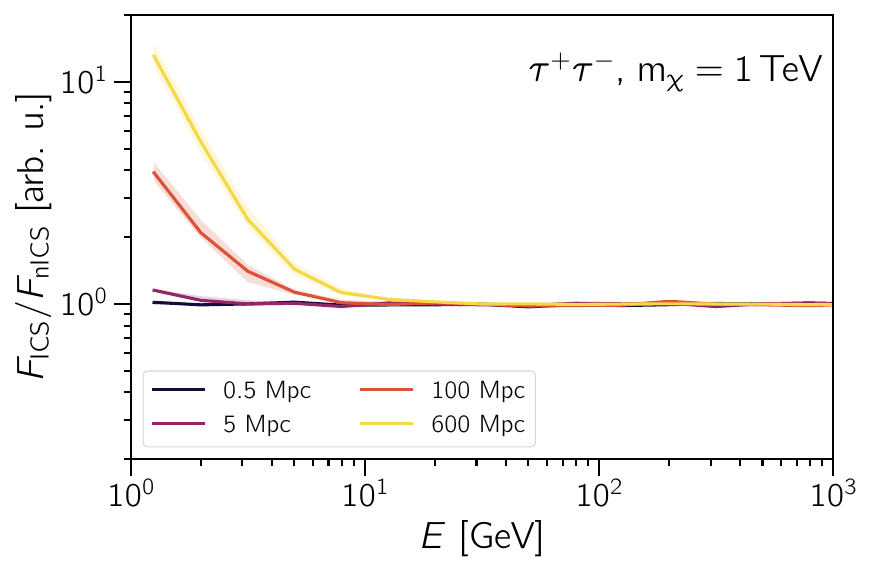}
\includegraphics[width=0.495\textwidth]{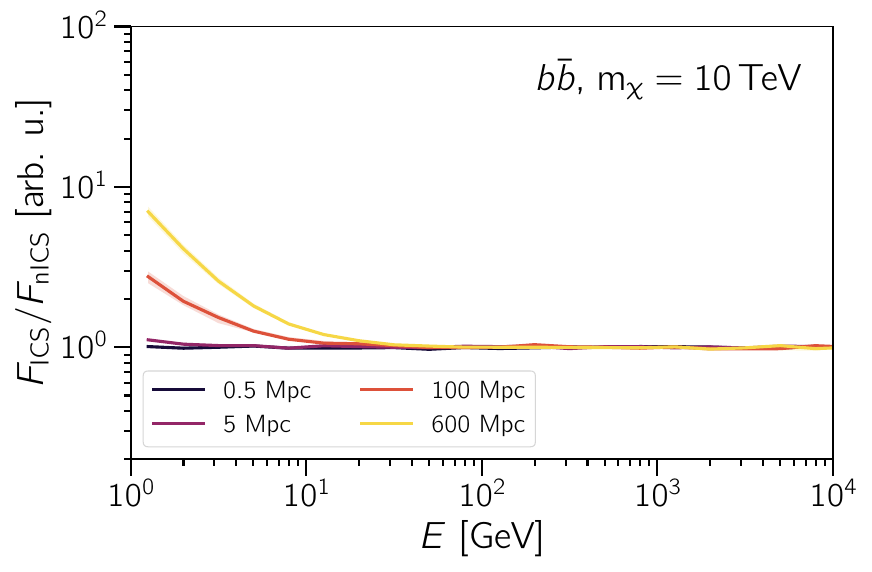}
\includegraphics[width=0.495\textwidth]{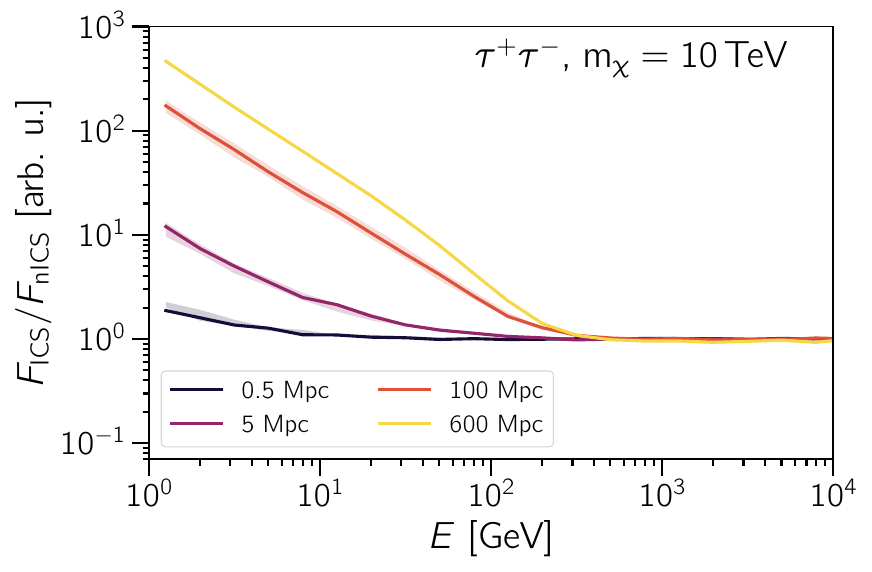}
\includegraphics[width=0.495\textwidth]{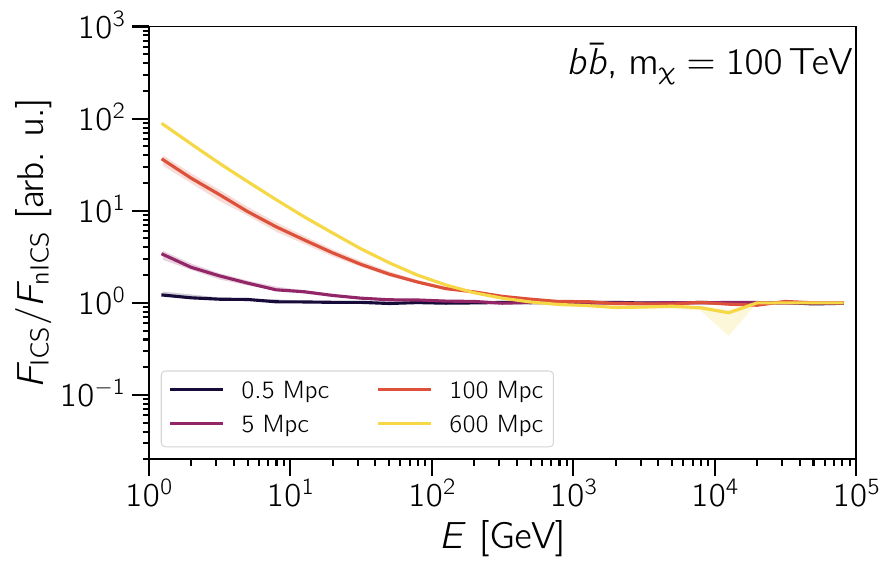}
\includegraphics[width=0.495\textwidth]{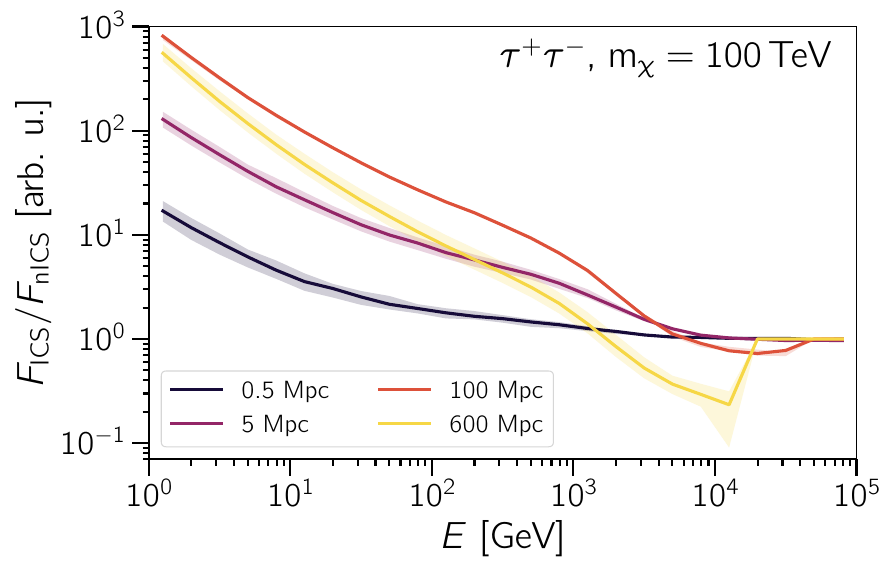}
\caption{Ratio between the expected gamma-ray fluxes at Earth including inverse Compton scattering (ICS) and neglecting it (nICS), $F_{\mathrm{ICS}}/F_{\mathrm{nICS}}$, as a function of energy (in GeV) for the $b\bar{b}$ (left column) and $\tau^{+}\tau^{-}$ (right column) annihilation channels. The panels correspond to DM masses of 1~TeV (upper row), 10~TeV (middle row), and 100~TeV (lower row). The black, purple, red, and yellow curves correspond to source distances of 0.5, 5, 100, and 600~Mpc, respectively. The shaded bands represent the combined uncertainty arising from the lower and upper Saldana-Lopez21 EBL models~\cite{Saldana_Lopez_2021} together with the Monte Carlo statistical uncertainty.}
\label{fig: F_ICS/F_nICS}
\end{figure*}

The results are shown in Figures~\ref{fig: Spectrum} and \ref{fig: F_ICS/F_nICS}. Figure~\ref{fig: Spectrum} displays the intrinsic spectrum of annihilation and the resulting dimensionless flux after propagation over various distances for the inverse Compton scattering and pair production interactions mentioned in Subsection~\ref{subsec: CRPropa}. Figure~\ref{fig: F_ICS/F_nICS} shows the ratio between fluxes calculated with inverse Compton scattering and in the absence of inverse Compton scattering for the same scenarios.

The enhancement observed at low energies in Figures~\ref{fig: Spectrum} and \ref{fig: F_ICS/F_nICS} reflects the gamma-ray regeneration mechanism discussed in Section~\ref{sec: Propagation}. As energy is progressively redistributed through the electromagnetic cascade, the low-energy flux increases while the highest-energy component is attenuated. This behaviour becomes more pronounced for larger propagation distances and higher dark matter masses, where cascade development is more efficient.

\subsection{The case of Perseus galaxy cluster}
\label{subsec: Perseus}

After analysing generic distances, we now focus on the well-known Perseus galaxy cluster, located at redshift $z \sim 0.017$~\citep{kang2024_perseus}. Perseus was selected as a benchmark target because it combines a large dark matter content with a sufficiently long propagation distance for electromagnetic cascade effects to become relevant. Furthermore, it has been extensively observed by MAGIC and constitutes one of the most promising nearby galaxy clusters for indirect dark matter searches \cite{S_nchez_Conde_2011}.

\begin{figure}[htb]
\centering
\includegraphics[width=0.6\textwidth]{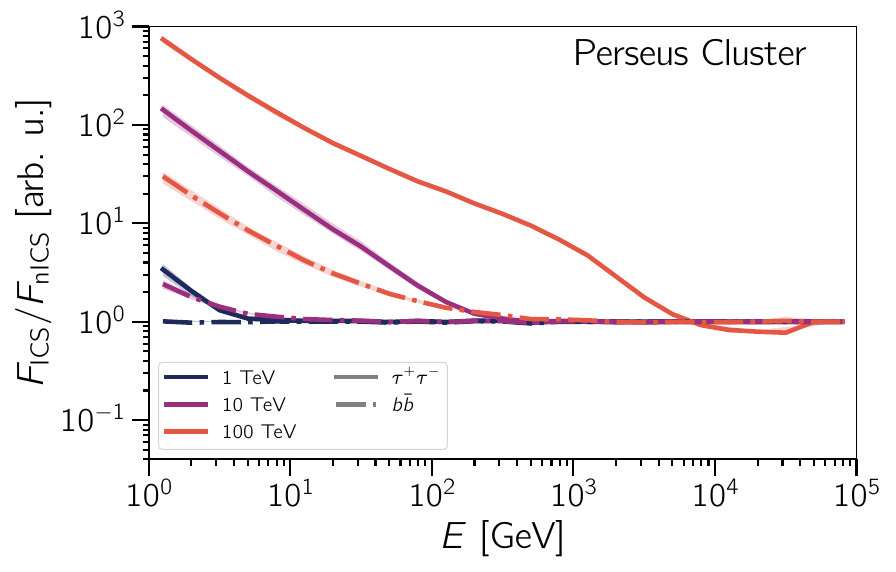}
\caption{Flux ratio ($F_\text{ICS} / F_\text{nICS}$) for the Perseus Cluster, as a function of the energy, for the $b\bar{b}$ and $\tau^{+}\tau^{-}$ channels. Black, purple and red lines correspond to DM masses of 1, 10, and 100~TeV, respectively. The shaded bands represent the uncertainty associated with the lower and upper Saldana-Lopez21 EBL models~\cite{Saldana_Lopez_2021}, combined with the Monte Carlo statistical uncertainty.}
\label{fig: Cluster}
\end{figure}

Figure~\ref{fig: Cluster} shows the flux ratio with (ICS) and without (nICS) inverse Compton scattering, labelled ($F_\text{ICS} / F_\text{nICS}$), for Perseus cluster and various DM masses and annihilation channels. While in the previous subsection we appreciate the relation between the flux ratio and the distance that gamma rays travel, in this case the results show how the WIMP mass affects this ratio, given that the distance is fixed. This figure seems to show a behaviour similar to the 100~Mpc cases in Fig.~\ref{fig: F_ICS/F_nICS}, since this is approximately the distance to the cluster.

\subsection{Impact on DM limits}
\label{subsec: DM limits}

\begin{figure*}[ht!]
\centering
\includegraphics[width=0.495\textwidth]{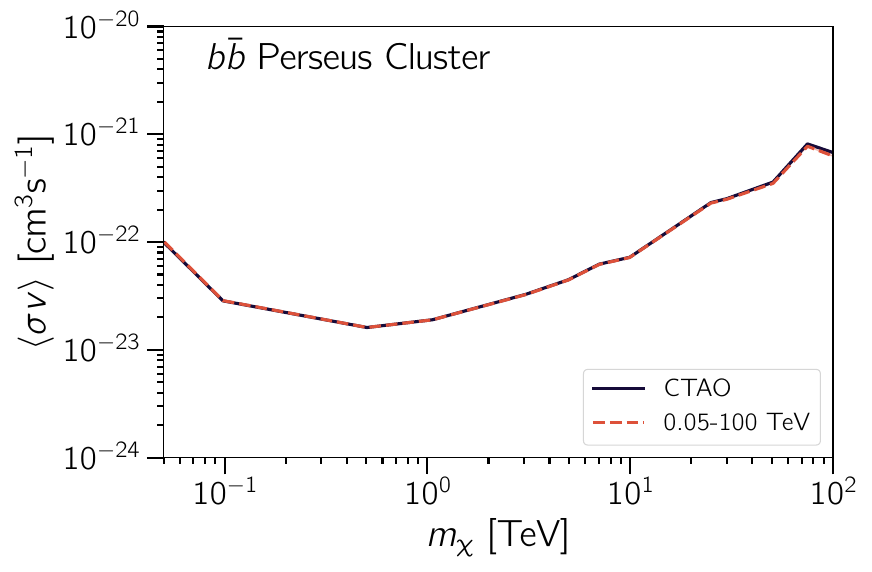}
\includegraphics[width=0.495\textwidth]{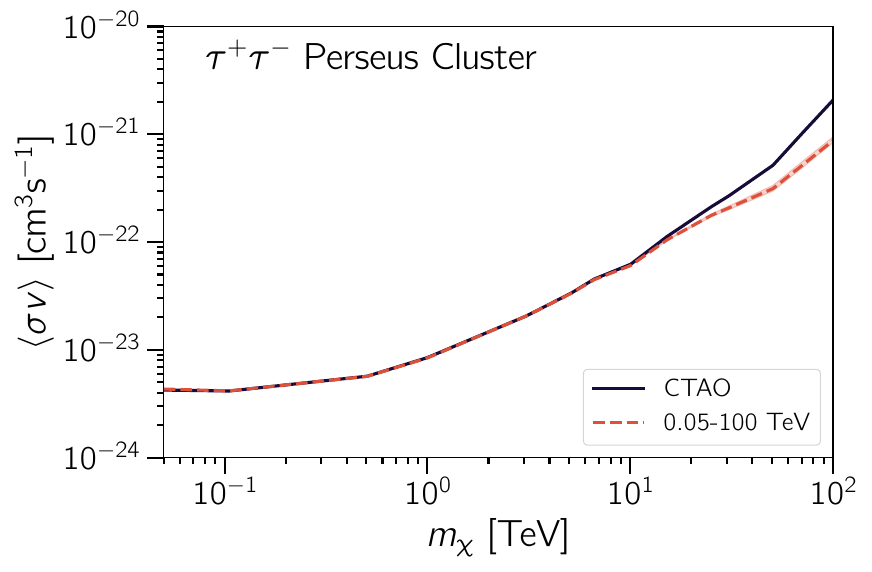}
\caption{Comparison of the exclusion limits on $\langle \sigma v \rangle$ obtained in this work (red dashed line, with the shaded band indicating the combined EBL~\cite{Saldana_Lopez_2021} and Monte Carlo uncertainties) and those from Ref.~\cite{Abe_2024} (solid black line).}
\label{fig: change upper}
\end{figure*}

To quantify the impact of inverse Compton scattering on DM exclusion limits, we integrate the normalised gamma-ray flux shown in Figure~\ref{fig: Cluster} over a range of WIMP masses $m_{\chi}$ in the Perseus cluster, comparing the results obtained with and without including the contribution from inverse Compton scattering. This procedure yields flux ratios that directly translate into modifications of the exclusion limits on $\langle \sigma v \rangle$, as illustrated in Figure~\ref{fig: change upper}. The results displayed in this figure were derived using as a reference an energy range (0.05-100 TeV) within the projected sensitivity of the CTAO from Ref.~\cite{Abe_2024}, and applying the factor corresponding to the difference between including and neglecting inverse Compton scattering. The obtained results suggest that a more detailed treatment of propagation does not lead to substantial changes in the main DM observable, namely the velocity-averaged annihilation cross section. While this is indeed the case, it strongly depends on the limits of integration over energy in Eq.~\ref{eq: ann and prop}. We thus examine the behaviour within specific energy intervals.

\begin{figure*}[ht!]
\centering
\includegraphics[width=0.495\textwidth]{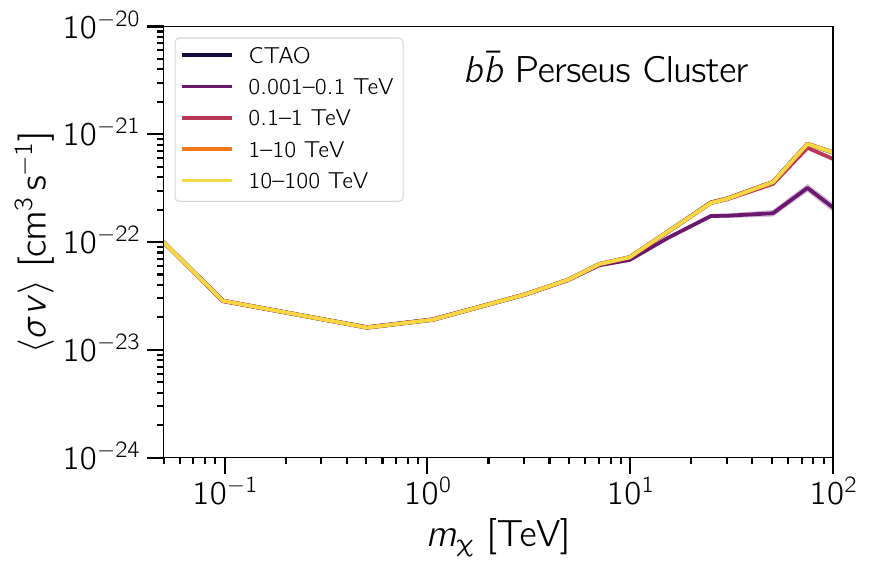}
\includegraphics[width=0.495\textwidth]{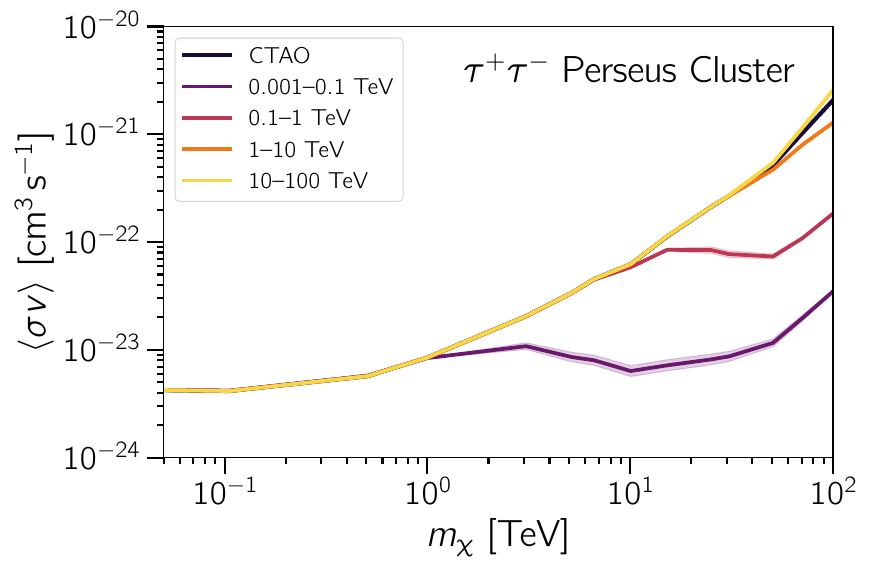}
\caption{Comparison of the exclusion limits on $\langle \sigma v \rangle$ from Ref.~\cite{Abe_2024} (solid black line), derived from the gamma-ray energy spectrum, and those obtained in this work for the integration ranges $0.001$--$0.1$~TeV (solid purple line), $0.1$--$1$~TeV (solid red line), $1$--$10$~TeV (solid orange line), and $10$--$100$~TeV (solid yellow line). The shaded bands represent the combined uncertainty associated with the lower and upper Saldana-Lopez21 EBL models~\cite{Saldana_Lopez_2021} and the Monte Carlo statistical uncertainty.}
\label{fig: all ranges}
\end{figure*}

To further explore this energy dependence, we repeat the comparison for different energy intervals. The resulting flux ratios and their influence on the limits are shown in Figure~\ref{fig: all ranges}.

This figure is instructive, as it illustrates how the parameter space evolves with the choice of an observational energy window. In particular, constraints derived from observatories sensitive to $\sim$~1–100~GeV energies (e.g., Fermi-LAT) show strong sensitivity to high DM masses for this specific cluster. This is because the low-energy emission is largely dominated by secondary photons produced through electromagnetic cascading rather than by the primary annihilation spectrum. By contrast, in the $\gtrsim 1 \;\text{TeV}$ range, typical of gamma rays detected by IACTs, the impact of propagation effects on the DM limits remains minimal, largely independent of the WIMP mass.

\section{Discussion}
\label{sec: Discussion}

The results presented in Section~\ref{sec: Results} highlight the crucial role of inverse Compton scattering in shaping the gamma-ray spectra originating from WIMP annihilation and observed at Earth. Across the range of WIMP masses and propagation distances studied, the inclusion of this process systematically redistributes energy from the highest-energy photons towards lower energies, leading to an enhanced low-energy flux and a corresponding depletion of the high-energy component.
As discussed in Section~\ref{sec: Propagation}, this behaviour originates from gamma-ray regeneration through electromagnetic cascading. High-energy photons undergoing pair production transfer their energy to secondary electrons and positrons, which subsequently upscatter background photons through inverse Compton scattering, partially re-injecting energy into the gamma-ray channel.

More specifically, for higher WIMP masses (e.g., $m_{\chi}\simeq100$~TeV) and propagation distances exceeding 100~Mpc, the impact becomes substantial, with flux ratios between simulations including and neglecting inverse Compton scattering reaching values as high as $10^{3}$ in the low-energy regime. This effect progressively weakens for lower masses and shorter propagation distances, where fewer cascade generations develop and the resulting spectra remain close to those obtained without inverse Compton scattering.

Some channels are more prone to the regeneration of photons via inverse Compton scattering. While hadronic channels such as $b\bar{b}$ exhibit more moderate deviations, typically up to one order of magnitude, leptonic channels like $\tau^{+}\tau^{-}$ display significantly larger differences, particularly for gamma rays arriving at Earth with energies below 1~TeV. These findings are consistent with the higher production rates of high-energy electrons and positrons produced in leptonic annihilation, which are responsible for the signal through the upscattering of background photons from the CMB and EBL.

In the specific case of the Perseus galaxy cluster, the results are similar to the generic scenarios at a similar distance. The flux ratios and modified exclusion limits suggest that these propagation effects are not merely theoretical subtleties, but can have measurable observational consequences. Notably, the shifts in the velocity-averaged DM cross-section constraints for the $\tau^{+}\tau^{-}$ channel approach factors of $\sim0.4$ for the highest masses considered, confirming that accurate modelling of gamma-ray propagation is essential when interpreting indirect DM detection limits.

Similarly, the interpretation of gamma-ray constraints for superheavy DM candidates with masses above $\sim 100 \; \text{TeV}$ must necessarily account for electromagnetic cascades initiated by their annihilation or decay products. On Galactic scales, interactions of the resulting high-energy photons and electrons with the interstellar radiation field (ISRF), as well as with the CMB and EBL are also important. These interactions reprocess the primary emission through inverse Compton scattering and pair production, softening the spectrum and modifying its observable shape and intensity~\cite{dimarco2024}. Consequently, existing gamma-ray bounds on such superheavy WIMPs may require revision as well. This is left for future work.

Overall, our study highlights the potentially important role of inverse Compton scattering of secondary electrons by backgrounds photons -- a phenomenon often neglected -- which can lead to systematic biases in both the predicted spectra and the derived constraints on DM properties, especially in the high-mass regime and for certain annihilation channels. 

Given the energies at which these discrepancies become important, gamma-ray observations in the 1--100~GeV energy range play a crucial role. At present, Fermi-LAT remains the only instrument covering this energy window. For more distant sources and higher DM masses, the effects discussed here may already be detectable with current IACTs.

\section{Conclusions and Outlook}
\label{sec: Conclusion}

In this work, we have developed a framework combining CosmiXs and CRPropa to consistently model both the intrinsic gamma-ray spectra from WIMP annihilation and their modification due to propagation effects. The main novelty is the inclusion of secondary photons produced via inverse Compton scattering generated by electrons and positrons produced through pair production. Our results demonstrate that accounting for this process significantly alters the predicted fluxes at Earth, with differences spanning up to three orders of magnitude depending on the considered energy range, WIMP mass, and annihilation channel.

For observational facilities such as Fermi-LAT and the upcoming CTAO, which probe precisely the energy ranges most affected by these interactions, our findings imply that inverse Compton scattering must be incorporated into any robust interpretation of gamma-ray data in the context of DM searches. In particular, the shifts in DM exclusion limits and cross-section estimates demonstrate that propagation effects can partially relax constraints previously considered stringent, potentially reopening portions of the WIMP parameter space.

Future work could extend this approach by including additional annihilation channels and by exploring the role of magnetic fields in shaping the secondary emission. Moreover, applying this framework to specific astrophysical targets beyond the one studied in this work could further test the universality of these findings across different environments.

In conclusion, our study highlights the importance of properly accounting for propagation effects in indirect DM searches. By showing how inverse Compton scattering can reshape the predicted DM-induced gamma-ray spectra and modify derived DM constraints, this work underscores that detailed modelling is not just a refinement but a necessity for accurate interpretation of future observations. Indeed, incorporating such effects will be key to obtaining more precise and reliable limits on WIMP properties using gamma-ray data collected by current instruments like Fermi-LAT and upcoming ones like CTAO.

\section*{Acknowledgements}

This work has been partially co-funded by the Spanish Ministry of Science, Innovation, and Universities through the MINERIA project (Grant ID: PID2024-157169OB-I00). IML gratefully acknowledges CIEMAT for supporting his predoctoral contract. RAB is supported by the Agence Nationale de la Recherche (ANR),
project ANR-23-CPJ1-0103-01. The early stages of this work were funded by the ``la Caixa'' Foundation (ID 100010434) and the European Union’s Horizon 2020 research and innovation program under the Marie Skłodowska-Curie grant agreement No 847648 (fellowship LCF/BQ/PI21/11830030). The work of MASC was supported by the grants PID2024-155874NB-C21 and CEX2020-001007-S, both funded by MCIN/AEI/10.13039/501100011033 and by ``ERDF A way of making Europe''. MASC also acknowledges the MultiDark Network, ref. RED2022-134411-T.

\bibliographystyle{JHEP}
\bibliography{bibliography.bib}

@ARTICLE{Bertone_2005,
       author = {{Bertone}, Gianfranco and {Hooper}, Dan and {Silk}, Joseph},
        title = "{{Particle dark matter: evidence, candidates and constraints}}",
      journal = {Phys. Rept.},
         year = 2005,
        month = jan,
       volume = {405},
       number = {5-6},
        pages = {279-390},
          doi = {10.1016/j.physrep.2004.08.031},
archivePrefix = {arXiv},
       eprint = {hep-ph/0404175},
 primaryClass = {hep-ph},
       adsurl = {https://ui.adsabs.harvard.edu/abs/2005PhR...405..279B}
}

@ARTICLE{Bertone_2018,
       author = {{Bertone}, Gianfranco and {Hooper}, Dan},
        title = "{History of dark matter}",
      journal = {Rev. Mod. Phys.},
         year = 2018,
        month = oct,
       volume = {90},
       number = {4},
          eid = {045002},
        pages = {045002},
          doi = {10.1103/RevModPhys.90.045002},
archivePrefix = {arXiv},
       eprint = {1605.04909},
 primaryClass = {astro-ph.CO},
       adsurl = {https://ui.adsabs.harvard.edu/abs/2018RvMP...90d5002B}
}

@ARTICLE{GONDOLO1991145,
    author = "Gondolo, Paolo and Gelmini, Graciela",
    title = "{Cosmic abundances of stable particles: Improved analysis}",
    reportNumber = "UCLA-90-TEP-68",
    doi = "10.1016/0550-3213(91)90438-4",
    journal = "Nucl. Phys. B",
    volume = "360",
    pages = "145--179",
    year = "1991"
}

@ARTICLE{Griest91,
    author = "Griest, Kim and Seckel, David",
    title = "{Three exceptions in the calculation of relic abundances}",
    reportNumber = "CFPA-TH-90-001A, BA-90-79",
    doi = "10.1103/PhysRevD.43.3191",
    journal = "Phys. Rev. D",
    volume = "43",
    pages = "3191--3203",
    year = "1991"
}

@ARTICLE{SREDNICKI1988693,
       author = {{Srednicki}, Mark and {Watkins}, Richard and {Olive}, Keith A.},
        title = "{Calculations of relic densities in the early universe}",
      journal = {Nucl. Phys. B},
         year = 1988,
        month = dec,
       volume = {310},
       number = {3-4},
        pages = {693-713},
          doi = {10.1016/0550-3213(88)90099-5},
       adsurl = {https://ui.adsabs.harvard.edu/abs/1988NuPhB.310..693S}
}

@ARTICLE{STEIGMAN1985375,
    author = {{Steigman}, Gary and {Turner}, Michael S.},
    title = "{{Cosmological constraints on the properties of weakly interacting massive particles}}",
    journal = {Nucl. Phys. B},
    year = 1985,
    month = jan,
    volume = {253},
    pages = {375-386},
    doi = {10.1016/0550-3213(85)90537-1},
    adsurl = {https://ui.adsabs.harvard.edu/abs/1985NuPhB.253..375S}
}

@ARTICLE{HABER198575,
       author = {{Haber}, H.~E. and {Kane}, G.~L.},
        title = "{{The search for supersymmetry: Probing physics beyond the standard model}}",
      journal = {Phys. Rept.},
         year = 1985,
        month = jan,
       volume = {117},
       number = {2-4},
        pages = {75-263},
          doi = {10.1016/0370-1573(85)90051-1},
       adsurl = {https://ui.adsabs.harvard.edu/abs/1985PhR...117...75H}
}

@ARTICLE{Arcadi_2018,
   title={The waning of the WIMP? A review of models, searches, and constraints},
   volume={78},
   ISSN={1434-6052},
   url={http://dx.doi.org/10.1140/epjc/s10052-018-5662-y},
   DOI={10.1140/epjc/s10052-018-5662-y},
   number={3},
   journal={The European Physical Journal C},
   publisher={Springer Science and Business Media LLC},
   author={Arcadi, Giorgio and Dutra, Maíra and Ghosh, Pradipta and Lindner, Manfred and Mambrini, Yann and Pierre, Mathias and Profumo, Stefano and Queiroz, Farinaldo S.},
   year={2018},
   month=Mar }

@ARTICLE{Roszkowski_2018,
       author = {{Roszkowski}, Leszek and {Sessolo}, Enrico Maria and {Trojanowski}, Sebastian},
        title = "{{WIMP dark matter candidates and searches{\textemdash}current status and future prospects}}",
      journal = {Rept. Prog. Phys.},
         year = 2018,
        month = jun,
       volume = {81},
       number = {6},
          eid = {066201},
        pages = {066201},
          doi = {10.1088/1361-6633/aab913},
archivePrefix = {arXiv},
       eprint = {1707.06277},
 primaryClass = {hep-ph},
       adsurl = {https://ui.adsabs.harvard.edu/abs/2018RPPh...81f6201R}
}

@ARTICLE{boveia2018a,
       author = {{Boveia}, Antonio and {Doglioni}, Caterina},
        title = "{Dark Matter Searches at Colliders}",
      journal = {Ann. Rev. of Nucl. Part. Sci.},
         year = 2018,
        month = oct,
       volume = {68},
        pages = {429-459},
          doi = {10.1146/annurev-nucl-101917-021008},
archivePrefix = {arXiv},
       eprint = {1810.12238},
 primaryClass = {hep-ex},
       adsurl = {https://ui.adsabs.harvard.edu/abs/2018ARNPS..68..429B}
}

@ARTICLE{lee2019a,
    author = {{Lee}, Lawrence and {Ohm}, Christian and {Soffer}, Abner and {Yu}, Tien-Tien},
    title = "{Collider searches for long-lived particles beyond the Standard Model}",
    journal = {Prog. Part. Nucl. Phys.},
    year = 2019,
    month = may,
    volume = {106},
    pages = {210-255},
    doi = {10.1016/j.ppnp.2019.02.006},
    archivePrefix = {arXiv},
    eprint = {1810.12602},
    primaryClass = {hep-ph},
    adsurl = {https://ui.adsabs.harvard.edu/abs/2019PrPNP.106..210L}
}

@ARTICLE{misiaszek2024a,
    author = "{Misiaszek}, Marcin and {Rossi}, Nicola",
    title = "{Direct Detection of Dark Matter: A Critical Review}",
    journal = "Symmetry",
    year = "2024",
    month = "February",
    volume = "16",
    number = "2",
    eid = "201",
    pages = "201",
    doi = "10.3390/sym16020201",
    archivePrefix = "arXiv",
    eprint = "2310.20472",
    primaryClass = "hep-ph",
    adsurl = "https://ui.adsabs.harvard.edu/abs/2024Symm...16..201M"
}

@ARTICLE{Gaskins_2016,
       author = {{Gaskins}, Jennifer M.},
        title = "{A review of indirect searches for particle dark matter}",
      journal = {Contemp.\ Phys.},
         year = 2016,
        month = oct,
       volume = {57},
       number = "4",
        pages = "496--525",
          doi = {10.1080/00107514.2016.1175160},
archivePrefix = {arXiv},
       eprint = {1604.00014},
 primaryClass = {astro-ph.HE},
       adsurl = {https://ui.adsabs.harvard.edu/abs/2016ConPh..57..496G}
}

@inproceedings{Leane:2020liq,
    author = "Leane, Rebecca K.",
    title = "{Indirect Detection of Dark Matter in the Galaxy}",
    booktitle = "{3rd World Summit on Exploring the Dark Side of the Universe}",
    address = {Guadeloupe Islands},
    publisher = {University of Kansas Libraries},
    eprint = "2006.00513",
    archivePrefix = "arXiv",
    primaryClass = "hep-ph",
    reportNumber = "MIT-CTP/5199",
    pages = "203--228",
    year = "2020"
}

@article{Ajello_2016,
    author = "Ajello, M. and others",
    collaboration = "Fermi-LAT",
    title = "{Fermi-LAT Observations of High-Energy $\ensuremath{\gamma}$-Ray Emission Toward the Galactic Center}",
    eprint = "1511.02938",
    archivePrefix = "arXiv",
    primaryClass = "astro-ph.HE",
    doi = "10.3847/0004-637X/819/1/44",
    journal = "Astrophys. J.",
    volume = "819",
    number = "1",
    pages = "44",
    year = "2016"
}

@article{Ackermann_2017,
    author = "Ackermann, M. and others",
    collaboration = "Fermi-LAT",
    title = "{The Fermi Galactic Center GeV Excess and Implications for Dark Matter}",
    eprint = "1704.03910",
    archivePrefix = "arXiv",
    primaryClass = "astro-ph.HE",
    doi = "10.3847/1538-4357/aa6cab",
    journal = "Astrophys. J.",
    volume = "840",
    number = "1",
    pages = "43",
    year = "2017"
}

@article{DiMauro2021mar,
    author = "Di Mauro, Mattia",
    title = "{Characteristics of the Galactic Center excess measured with 11 years of $Fermi$-LAT data}",
    eprint = "2101.04694",
    archivePrefix = "arXiv",
    primaryClass = "astro-ph.HE",
    doi = "10.1103/PhysRevD.103.063029",
    journal = "Phys. Rev. D",
    volume = "103",
    number = "6",
    pages = "063029",
    year = "2021"
}

@article{Zuriaga-Puig_2023,
    author = "Zuriaga-Puig, Jaume and Gammaldi, Viviana and Gaggero, Daniele and Lacroix, Thomas and S{\'a}nchez-Conde, Miguel {\'A}ngel",
    title = "{Multi-TeV dark matter density in the inner Milky Way halo: spectral and dynamical constraints}",
    eprint = "2307.06823",
    archivePrefix = "arXiv",
    primaryClass = "astro-ph.HE",
    doi = "10.1088/1475-7516/2023/11/063",
    journal = "JCAP",
    volume = "11",
    pages = "063",
    year = "2023"
}

@article{Ackerman2015nov,
  title = "{Searching for Dark Matter Annihilation from Milky Way Dwarf Spheroidal Galaxies with Six Years of Fermi Large Area Telescope Data}",
  author = {Ackermann, M. and Albert, A. and Anderson, B. and Atwood, W. B. and Baldini, L. and Barbiellini, G. and Bastieri, D. and Bechtol, K. and Bellazzini, R. and Bissaldi, E. and Blandford, R. D. and Bloom, E. D. and Bonino, R. and Bottacini, E. and Brandt, T. J. and Bregeon, J. and Bruel, P. and Buehler, R. and Caliandro, G. A. and Cameron, R. A. and Caputo, R. and Caragiulo, M. and Caraveo, P. A. and Cecchi, C. and Charles, E. and Chekhtman, A. and Chiang, J. and Chiaro, G. and Ciprini, S. and Claus, R. and Cohen-Tanugi, J. and Conrad, J. and Cuoco, A. and Cutini, S. and D'Ammando, F. and de Angelis, A. and de Palma, F. and Desiante, R. and Digel, S. W. and Di Venere, L. and Drell, P. S. and Drlica-Wagner, A. and Essig, R. and Favuzzi, C. and Fegan, S. J. and Ferrara, E. C. and Focke, W. B. and Franckowiak, A. and Fukazawa, Y. and Funk, S. and Fusco, P. and Gargano, F. and Gasparrini, D. and Giglietto, N. and Giordano, F. and Giroletti, M. and Glanzman, T. and Godfrey, G. and Gomez-Vargas, G. A. and Grenier, I. A. and Guiriec, S. and Gustafsson, M. and Hays, E. and Hewitt, J. W. and Horan, D. and Jogler, T. and J\'ohannesson, G. and Kuss, M. and Larsson, S. and Latronico, L. and Li, J. and Li, L. and Llena Garde, M. and Longo, F. and Loparco, F. and Lubrano, P. and Malyshev, D. and Mayer, M. and Mazziotta, M. N. and McEnery, J. E. and Meyer, M. and Michelson, P. F. and Mizuno, T. and Moiseev, A. A. and Monzani, M. E. and Morselli, A. and Murgia, S. and Nuss, E. and Ohsugi, T. and Orienti, M. and Orlando, E. and Ormes, J. F. and Paneque, D. and Perkins, J. S. and Pesce-Rollins, M. and Piron, F. and Pivato, G. and Porter, T. A. and Rain\`o, S. and Rando, R. and Razzano, M. and Reimer, A. and Reimer, O. and Ritz, S. and S\'anchez-Conde, M. and Schulz, A. and Sehgal, N. and Sgr\`o, C. and Siskind, E. J. and Spada, F. and Spandre, G. and Spinelli, P. and Strigari, L. and Tajima, H. and Takahashi, H. and Thayer, J. B. and Tibaldo, L. and Torres, D. F. and Troja, E. and Vianello, G. and Werner, M. and Winer, B. L. and Wood, K. S. and Wood, M. and Zaharijas, G. and Zimmer, S.},
  collaboration = {Fermi-LAT},
  journal = {Phys. Rev. Lett.},
  volume = {115},
  issue = {23},
  pages = {231301},
  numpages = {8},
  year = {2015},
  month = {Nov},
  publisher = {American Physical Society},
  doi = {10.1103/PhysRevLett.115.231301},
  url = {https://link.aps.org/doi/10.1103/PhysRevLett.115.231301}
}

@article{armand2021,
    author = "Abdalla, Hassan and others",
    collaboration = "Hess, HAWC, VERITAS, MAGIC, H.E.S.S., Fermi-LAT",
    title = "{Combined dark matter searches towards dwarf spheroidal galaxies with Fermi-LAT, HAWC, H.E.S.S., MAGIC, and VERITAS}",
    eprint = "2108.13646",
    archivePrefix = "arXiv",
    primaryClass = "hep-ex",
    doi = "10.22323/1.395.0528",
    journal = "PoS",
    volume = "ICRC2021",
    pages = "528",
    year = "2021"
}

@article{Gammaldi2021,
    author = "Gammaldi, Viviana and P{\'e}rez-Romero, Judit and Coronado-Bl{\'a}zquez, Javier and Di Mauro, Mattia and Karukes, Ekaterina and S{\'a}nchez-Conde, Miguel Angel and Salucci., Paolo",
    collaboration = "Fermi-LAT",
    title = "{Dark Matter search in dwarf irregular galaxies with the Fermi Large Area Telescope}",
    eprint = "2109.11291",
    archivePrefix = "arXiv",
    primaryClass = "astro-ph.CO",
    doi = "10.22323/1.395.0509",
    journal = "PoS",
    volume = "ICRC2021",
    pages = "509",
    year = "2021"
}

@article{mcdaniel2023,
  title = "{Legacy analysis of dark matter annihilation from the Milky Way dwarf spheroidal galaxies with 14 years of $\mathit{Fermi}$-LAT data}",
  author = {McDaniel, Alex and Ajello, Marco and Karwin, Christopher M. and Di Mauro, Mattia and Drlica-Wagner, Alex and S\'anchez-Conde, Miguel A.},
  journal = {Phys. Rev. D},
  volume = {109},
  issue = {6},
  pages = {063024},
  numpages = {20},
  year = {2024},
  month = {Mar},
  publisher = {American Physical Society},
  doi = {10.1103/PhysRevD.109.063024},
      eprint={2311.04982},
      archivePrefix={arXiv},
      primaryClass={astro-ph.HE},
      url={https://arxiv.org/abs/2311.04982}
}

@article{fernandezsuarez2025,
      title="{A Search for Dark Matter Annihilation in Stellar Streams with the Fermi-LAT}", 
      author={Cristina Fernández-Suarez and Miguel A. Sánchez-Conde},
      journal={JCAP},
      month={\sep},
      volume={2025},
      number={09},
      pages={003},
      doi={10.1088/1475-7516/2025/09/003},
      year={2025},
      eprint={2502.15656},
      archivePrefix={arXiv},
      primaryClass={astro-ph.HE},
      url={https://arxiv.org/abs/2502.15656} 
}

@article{Ackermann_2015oct,
    author = "Ackermann, M. and others",
    collaboration = "Fermi-LAT",
    title = "{Search for extended gamma-ray emission from the Virgo galaxy cluster with Fermi-LAT}",
    eprint = "1510.00004",
    archivePrefix = "arXiv",
    primaryClass = "astro-ph.HE",
    doi = "10.1088/0004-637X/812/2/159",
    journal = "Astrophys. J.",
    volume = "812",
    number = "2",
    pages = "159",
    year = "2015"
}

@article{DiMauro2023,
    author = "Di Mauro, Mattia and P{\'e}rez-Romero, Judit and S{\'a}nchez-Conde, Miguel A. and Fornengo, Nicolao",
    title = "{Constraining the dark matter contribution of {\ensuremath{\gamma}} rays in clusters of galaxies using Fermi-LAT data}",
    eprint = "2303.16930",
    archivePrefix = "arXiv",
    primaryClass = "astro-ph.HE",
    doi = "10.1103/PhysRevD.107.083030",
    journal = "Phys. Rev. D",
    volume = "107",
    number = "8",
    pages = "083030",
    year = "2023"
}

@article{Ackermann_2012,
    author = "Ackermann, M. and others",
    collaboration = "Fermi-LAT",
    title = "{Search for Dark Matter Satellites using the Fermi-LAT}",
    eprint = "1201.2691",
    archivePrefix = "arXiv",
    primaryClass = "astro-ph.HE",
    reportNumber = "SLAC-PUB-14974",
    doi = "10.1088/0004-637X/747/2/121",
    journal = "Astrophys. J.",
    volume = "747",
    pages = "121",
    year = "2012"
}

@article{Coronado-Blazquez2019jul,
    author = "Coronado-Blazquez, Javier and Sanchez-Conde, Miguel A. and Dominguez, Alberto and Aguirre-Santaella, Alejandra and Di Mauro, Mattia and Mirabal, Nestor and Nieto, Daniel and Charles, Eric",
    title = "{Unidentified Gamma-ray Sources as Targets for Indirect Dark Matter Detection with the Fermi-Large Area Telescope}",
    eprint = "1906.11896",
    archivePrefix = "arXiv",
    primaryClass = "astro-ph.HE",
    doi = "10.1088/1475-7516/2019/07/020",
    journal = "JCAP",
    volume = "07",
    pages = "020",
    year = "2019"
}

@article{Coronado-Blazquez2019nov,
author = {Coronado-Blázquez, Javier and Sánchez-Conde, Miguel A. and Mauro, Mattia Di and Aguirre-Santaella, Alejandra and Ciucă, Ioana and Domínguez, Alberto and Kawata, Daisuke and Mirabal, Néstor},
title = {Spectral and spatial analysis of the dark matter subhalo candidates among Fermi Large Area Telescope unidentified sources},
    collaboration = "Fermi-LAT",
    eprint = "2204.00267",
    archivePrefix = "arXiv",
    primaryClass = "astro-ph.HE",
    doi = "10.1103/PhysRevD.105.083006",
year = {2019},
month = {nov},
publisher = {},
volume = {2019},
number = {11},
pages = {045},
url = {https://doi.org/10.1088/1475-7516/2019/11/045},


journal = {JCAP}
}

@article{Coronado-Blazquez2022,
  title = {Spatial extension of dark subhalos as seen by Fermi-LAT and the implications for WIMP constraints},
  author = {Coronado-Bl\'azquez, Javier and S\'anchez-Conde, Miguel A. and P\'erez-Romero, Judit and Aguirre-Santaella, Alejandra},
  collaboration = {Fermi-LAT},
  journal = {Phys. Rev. D},
  volume = {105},
  issue = {8},
  pages = {083006},
  numpages = {22},
  year = {2022},
  month = {Apr},
  publisher = {American Physical Society},
    eprint = "2204.00267",
    archivePrefix = "arXiv",
    primaryClass = "astro-ph.HE",
    doi = "10.1103/PhysRevD.105.083006",
  url = {https://link.aps.org/doi/10.1103/PhysRevD.105.083006}
}

@ARTICLE{Atwood_2009,
    author = "Atwood, W. B. and others",
    collaboration = "Fermi-LAT",
    title = "{The Large Area Telescope on the Fermi Gamma-ray Space Telescope Mission}",
    eprint = "0902.1089",
    archivePrefix = "arXiv",
    primaryClass = "astro-ph.IM",
    reportNumber = "SLAC-PUB-13620",
    doi = "10.1088/0004-637X/697/2/1071",
    journal = "Astrophys. J.",
    volume = "697",
    pages = "1071--1102",
    year = "2009"
}

@ARTICLE{Conrad_2015,
    author = "Conrad, Jan and Cohen-Tanugi, Johann and Strigari, Louis E.",
    title = "{WIMP searches with gamma rays in the Fermi era: challenges, methods and results}",
    eprint = "1503.06348",
    archivePrefix = "arXiv",
    primaryClass = "astro-ph.CO",
    doi = "10.1134/S1063776115130099",
    journal = "J. Exp. Theor. Phys.",
    volume = "121",
    number = "6",
    pages = "1104--1135",
    year = "2015"
}

@ARTICLE{Charles_2016,
    author = "Charles, E. and others",
    collaboration = "Fermi-LAT",
    title = "{Sensitivity Projections for Dark Matter Searches with the Fermi Large Area Telescope}",
    eprint = "1605.02016",
    archivePrefix = "arXiv",
    primaryClass = "astro-ph.HE",
    reportNumber = "FERMILAB-PUB-16-179-AE",
    doi = "10.1016/j.physrep.2016.05.001",
    journal = "Phys. Rept.",
    volume = "636",
    pages = "1--46",
    year = "2016"
}

@article{WEEKES2002221,
    author = "Weekes, T. C. and others",
    title = "{VERITAS: The Very energetic radiation imaging telescope array system}",
    eprint = "astro-ph/0108478",
    archivePrefix = "arXiv",
    doi = "10.1016/S0927-6505(01)00152-9",
    journal = "Astropart. Phys.",
    volume = "17",
    pages = "221--243",
    year = "2002"
}

@article{LORENZ2004339,
title = {Status of the 17 m MAGIC telescope},
journal = {New Astron. Rev.},
volume = {48},
number = {5},
pages = {339-344},
year = {2004},
note = {2nd VERITAS Symposium on the Astrophysics of Extragalactic Sources},
issn = {1387-6473},
doi = {https://doi.org/10.1016/j.newar.2003.12.059},
url = {https://www.sciencedirect.com/science/article/pii/S1387647303003579},
author = {Eckart Lorenz}
}

@article{HINTON2004331,
title = {The status of the HESS project},
journal = {New Astron. Rev.},
volume = {48},
number = {5},
pages = {331-337},
year = {2004},
note = {2nd VERITAS Symposium on the Astrophysics of Extragalactic Sources},
issn = {1387-6473},
doi = {https://doi.org/10.1016/j.newar.2003.12.004},
url = {https://www.sciencedirect.com/science/article/pii/S1387647303003555},
author = {J.A Hinton}
}

@article{Abeysekara_2017,
    author = "Abeysekara, A. U. and others",
    collaboration = "HAWC",
    title = "{The 2HWC HAWC Observatory Gamma Ray Catalog}",
    eprint = "1702.02992",
    archivePrefix = "arXiv",
    primaryClass = "astro-ph.HE",
    doi = "10.3847/1538-4357/aa7556",
    journal = "Astrophys. J.",
    volume = "843",
    number = "1",
    pages = "40",
    year = "2017"
}

@book{Acharya2018,
    author = "Acharya, B. S. and others",
    collaboration = "CTA",
    title = "{Science with the Cherenkov Telescope Array}",
    eprint = "1709.07997",
    archivePrefix = "arXiv",
    primaryClass = "astro-ph.IM",
    doi = "10.1142/10986",
    isbn = "978-981-327-008-4",
    publisher = "WSP",
    month = "11",
    year = "2018"
}

@book{Mukherjee_2024,
  author = {Mukherjee, Reshmi and Zanin, Roberta},
  title = {Advances in Very High Energy Astrophysics: The Science Program of the Third Generation IACTs for Exploring Cosmic Gamma Rays},
  publisher = {World Scientific},
  address = {Singapore},
  year = {2024},
  pages = {492},
  doi = {10.1142/11141}
}

@article{Abreu_2025swgo,
  author = {Abreu, P. and others},
  collaboration = {SWGO},
  title = "{Science Prospects for the Southern Wide-field Gamma-ray Observatory: SWGO}",
  eprint = {2506.01786},
  archivePrefix = {arXiv},
  primaryClass = {astro-ph.HE},
  year = {2025},
  reportNumber = {SWGO Collaboration},
  note = {arXiv:2506.01786}
}

@article{Ma_2022,
  author = {Ma, Xin-Hua and others},
  collaboration = {LHAASO},
  title = {LHAASO Instruments and Detector Technology},
  journal = {Chin.\ Phys.\ C},
  volume = {46},
  number = {3},
  pages = {030001},
  year = {2022},
  doi = {10.1088/1674-1137/ac3fa6}
}

@ARTICLE{Morselli2023,
    author = "Morselli, Aldo",
    collaboration = "CTA",
    title = "{Search for dark matter with IACTs and the Cherenkov Telescope Array}",
    eprint = "2302.11318",
    archivePrefix = "arXiv",
    primaryClass = "astro-ph.HE",
    doi = "10.1088/1742-6596/2429/1/012019",
    journal = "J. Phys. Conf. Ser.",
    volume = "2429",
    number = "1",
    pages = "012019",
    year = "2023"
}

@ARTICLE{Salati2023,
    author = "Salati, Pierre",
    title = "{What charged cosmic rays tell us on dark matter}",
    eprint = "2210.07166",
    archivePrefix = "arXiv",
    primaryClass = "astro-ph.HE",
    reportNumber = "LAPTH-Conf-065/22",
    doi = "10.21468/SciPostPhysProc.12.067",
    journal = "SciPost Phys. Proc.",
    volume = "12",
    pages = "067",
    year = "2023"
}

@ARTICLE{Abe_2020,
    author = "Abe, K. and others",
    collaboration = "Super-Kamiokande",
    title = "{Indirect search for dark matter from the Galactic Center and halo with the Super-Kamiokande detector}",
    eprint = "2005.05109",
    archivePrefix = "arXiv",
    primaryClass = "hep-ex",
    doi = "10.1103/PhysRevD.102.072002",
    journal = "Phys. Rev. D",
    volume = "102",
    number = "7",
    pages = "072002",
    year = "2020"
}

@ARTICLE{Albert_2020,
  author = {{ANTARES Collaboration} and {IceCube Collaboration}},
  title = {Combined search for neutrinos from dark matter self-annihilation in the Galactic Center with ANTARES and IceCube},
  journal = {Physical Review D},
  year = {2020},
  volume = {102},
  number = {8},
  eid = {082002},
  pages = {082002},
  doi = {10.1103/PhysRevD.102.082002},
  archivePrefix = {arXiv},
  eprint = {2003.06614},
  primaryClass = {astro-ph.HE}
}

@ARTICLE{Aiello_2025,
  author = {{Aiello}, S. and others },
  collaboration = {{KM3NeT}},
  title = {{First searches for dark matter with the KM3NeT neutrino telescopes}},
  journal = {JCAP},
  year = {2025},
  month = mar,
  volume = {2025},
  number = {3},
  eid = {058},
  pages = {058},
  doi = {10.1088/1475-7516/2025/03/058},
  archivePrefix = {arXiv},
  eprint = {2411.10092},
  primaryClass = {astro-ph.HE},
  adsurl = {https://ui.adsabs.harvard.edu/abs/2025JCAP...03..058A}
}

@ARTICLE{Archambault_2017,
   title={{Dark matter constraints from a joint analysis of dwarf Spheroidal galaxy observations with VERITAS}},
   volume={95},
   ISSN={2470-0029},
   url={http://dx.doi.org/10.1103/PhysRevD.95.082001},
   DOI={10.1103/physrevd.95.082001},
   number={8},
   journal={Physical Review D},
   publisher={American Physical Society (APS)},
   author={Archambault and others},
   year={2017},
   eprint = "1703.04937",
   archivePrefix = "arXiv",
   month=Apr }

@ARTICLE{Abdallah_2018,
   title={{Search for $\ensuremath{\gamma}$-Ray Line Signals from Dark Matter Annihilations in the Inner Galactic Halo from 10 Years of Observations with H.E.S.S.}},
   collaboration = "HESS",
   volume={120},
   ISSN={1079-7114},
   url={http://dx.doi.org/10.1103/PhysRevLett.120.201101},
   DOI={10.1103/physrevlett.120.201101},
   number={20},
   journal={Physical Review Letters},
   publisher={American Physical Society (APS)},
   author={Abdallah and others},
   year={2018},
   eprint = "1805.05741",
   archivePrefix = "arXiv",
   month=May }

@ARTICLE{Abe_2023,
    author = "Abe, H. and others",
    collaboration = "MAGIC",
    title = "{Search for Gamma-Ray Spectral Lines from Dark Matter Annihilation up to 100~TeV toward the Galactic Center with MAGIC}",
    eprint = "2212.10527",
    archivePrefix = "arXiv",
    primaryClass = "astro-ph.HE",
    reportNumber = "KEK-TH-2487, KEK-Cosmo-0307",
    doi = "10.1103/PhysRevLett.130.061002",
    journal = "Phys. Rev. Lett.",
    volume = "130",
    number = "6",
    pages = "061002",
    year = "2023"
}

@article{Albert_2018,
    author = "Albert, A. and others",
    collaboration = "HAWC",
    title = "{Dark Matter Limits From Dwarf Spheroidal Galaxies with The HAWC Gamma-Ray Observatory}",
    eprint = "1706.01277",
    archivePrefix = "arXiv",
    primaryClass = "astro-ph.HE",
    doi = "10.3847/1538-4357/aaa6d8",
    journal = "Astrophys. J.",
    volume = "853",
    number = "2",
    pages = "154",
    year = "2018"
}

@article{Cao_2022,
    author = "Cao, Zhen and others",
    collaboration = "LHAASO",
    title = "{Constraints on Heavy Decaying Dark Matter from 570~Days of LHAASO Observations}",
    eprint = "2210.15989",
    archivePrefix = "arXiv",
    primaryClass = "astro-ph.HE",
    doi = "10.1103/PhysRevLett.129.261103",
    journal = "Phys. Rev. Lett.",
    volume = "129",
    number = "26",
    pages = "261103",
    year = "2022"
}

@inproceedings{driver2021,
    author = "Driver, Simon P.",
    title = "{Measuring energy production in the Universe over all wavelengths and all time}",
    booktitle = "{IAU Symposium 355}: {The Realm of the Low Surface Brightness Universe}",
    eprint = "2102.12089",
    archivePrefix = "arXiv",
    primaryClass = "astro-ph.CO",
    month = "2",
    year = "2021"
}

@ARTICLE{mather1994,
       author = {{Mather}, J.~C. and {Cheng}, E.~S. and {Cottingham}, D.~A. and {Eplee}, Jr., R.~E. and {Fixsen}, D.~J. and {Hewagama}, T. and {Isaacman}, R.~B. and {Jensen}, K.~A. and {Meyer}, S.~S. and {Noerdlinger}, P.~D. and {Read}, S.~M. and {Rosen}, L.~P. and {Shafer}, R.~A. and {Wright}, E.~L. and {Bennett}, C.~L. and {Boggess}, N.~W. and {Hauser}, M.~G. and {Kelsall}, T. and {Moseley}, Jr., S.~H. and {Silverberg}, R.~F. and {Smoot}, G.~F. and {Weiss}, R. and {Wilkinson}, D.~T.},
        title = "{Measurement of the Cosmic Microwave Background Spectrum by the COBE FIRAS Instrument}",
      journal = {Astrophys.\ J.},
         year = 1994,
        month = jan,
       volume = {420},
        pages = {439},
          doi = {10.1086/173574},
       adsurl = {https://ui.adsabs.harvard.edu/abs/1994ApJ...420..439M}
}

@ARTICLE{franceschini2008,
    author = "Franceschini, Alberto and Rodighiero, Giulia",
    title = "{The extragalactic background light revisited and the cosmic photon-photon opacity}",
    eprint = "1705.10256",
    archivePrefix = "arXiv",
    primaryClass = "astro-ph.HE",
    doi = "10.1051/0004-6361/201629684",
    journal = "Astron. Astrophys.",
    volume = "603",
    pages = "A34",
    year = "2017"
}

@ARTICLE{fixsen2011,
    author = "Fixsen, D. J. and others",
    title = "{ARCADE 2 Measurement of the Extra-Galactic Sky Temperature at 3-90 GHz}",
    eprint = "0901.0555",
    archivePrefix = "arXiv",
    primaryClass = "astro-ph.CO",
    doi = "10.1088/0004-637X/734/1/5",
    journal = "Astrophys. J.",
    volume = "734",
    pages = "5",
    year = "2011"
}

@article{MuraseBeacom2012,
    author = "Murase, Kohta and Beacom, John F.",
    title = "{Constraining Very Heavy Dark Matter Using Diffuse Backgrounds of Neutrinos and Cascaded Gamma Rays}",
    eprint = "1206.2595",
    archivePrefix = "arXiv",
    primaryClass = "hep-ph",
    doi = "10.1088/1475-7516/2012/10/043",
    journal = "JCAP",
    volume = "10",
    pages = "043",
    year = "2012"
}

@article{Song:2023xdk,
    author = "Song, Deheng and Murase, Kohta and Kheirandish, Ali",
    title = "{Constraining decaying very heavy dark matter from galaxy clusters with 14 year Fermi-LAT data}",
    eprint = "2308.00589",
    archivePrefix = "arXiv",
    primaryClass = "astro-ph.HE",
    doi = "10.1088/1475-7516/2024/03/024",
    journal = "JCAP",
    volume = "03",
    pages = "024",
    year = "2024"
}

@ARTICLE{Arina2024,
    author = "Arina, Chiara and Di Mauro, Mattia and Fornengo, Nicolao and Heisig, Jan and Jueid, Adil and de Austri, Roberto Ruiz",
    title = "{CosmiXs: cosmic messenger spectra for indirect dark matter searches}",
    eprint = "2312.01153",
    archivePrefix = "arXiv",
    primaryClass = "astro-ph.HE",
    reportNumber = "TTK-23-32, CTPU-PTC-23-36",
    doi = "10.1088/1475-7516/2024/03/035",
    journal = "JCAP",
    volume = "03",
    pages = "035",
    year = "2024"
}

@ARTICLE{Batista_2016,
       author = {{Alves Batista}, Rafael and {Dundovic}, Andrej and {Erdmann}, Martin and {Kampert}, Karl-Heinz and {Kuempel}, Daniel and {M{\"u}ller}, Gero and {Sigl}, Guenter and {van Vliet}, Arjen and {Walz}, David and {Winchen}, Tobias},
        title = "{CRPropa 3{\textemdash}a public astrophysical simulation framework for propagating extraterrestrial ultra-high energy particles}",
      journal = {JCAP},
         year = 2016,
        month = may,
       volume = {05},
       number = {5},
          eid = {038},
        pages = {038},
          doi = {10.1088/1475-7516/2016/05/038},
archivePrefix = {arXiv},
       eprint = {1603.07142},
 primaryClass = {astro-ph.IM},
       adsurl = {https://ui.adsabs.harvard.edu/abs/2016JCAP...05..038A}
}

@ARTICLE{Alves_Batista_2022,
       author = {{Alves Batista}, Rafael and {Becker Tjus}, Julia and {D{\"o}rner}, Julien and {Dundovic}, Andrej and {Eichmann}, Bj{\"o}rn and {Frie}, Antonius and {Heiter}, Christopher and {Hoerbe}, Mario R. and {Kampert}, Karl-Heinz and {Merten}, Lukas and {M{\"u}ller}, Gero and {Reichherzer}, Patrick and {Saveliev}, Andrey and {Schlegel}, Leander and {Sigl}, G{\"u}nter and {van Vliet}, Arjen and {Winchen}, Tobias},
        title = "{CRPropa 3.2 - an advanced framework for high-energy particle propagation in extragalactic and galactic spaces}",
      journal = {JCAP},
         year = 2022,
        month = sep,
       volume = {2022},
       number = {9},
          eid = {035},
        pages = {035},
          doi = {10.1088/1475-7516/2022/09/035},
archivePrefix = {arXiv},
       eprint = {2208.00107},
 primaryClass = {astro-ph.HE},
       adsurl = {https://ui.adsabs.harvard.edu/abs/2022JCAP...09..035A}
}

@book{Dodelson:2003ft,
  author = {Dodelson, Scott and Schmidt, Fabian},
  title = {Modern Cosmology},
  publisher = {Academic Press},
  year = {2020}
}

@article{Steigman_2012,
    author = "Steigman, Gary and Dasgupta, Basudeb and Beacom, John F.",
    title = "{Precise Relic WIMP Abundance and its Impact on Searches for Dark Matter Annihilation}",
    eprint = "1204.3622",
    archivePrefix = "arXiv",
    primaryClass = "hep-ph",
    doi = "10.1103/PhysRevD.86.023506",
    journal = "Phys. Rev. D",
    volume = "86",
    pages = "023506",
    year = "2012"
}

@ARTICLE{Arina_2010,
       author = {{Arina}, Chiara and {Hambye}, Thomas and {Ibarra}, Alejandro and {Weniger}, Christoph},
        title = "{Intense gamma-ray lines from hidden vector dark matter decay}",
      journal = {JCAP},
         year = 2010,
        month = mar,
       volume = {2010},
       number = {3},
          eid = {024},
        pages = {024},
          doi = {10.1088/1475-7516/2010/03/024},
archivePrefix = {arXiv},
       eprint = {0912.4496},
 primaryClass = {hep-ph},
       adsurl = {https://ui.adsabs.harvard.edu/abs/2010JCAP...03..024A}
}

@ARTICLE{Acciari_2018,
       author = {{Acciari}, V.~A. and {Ansoldi}, S. and {Antonelli}, L.~A. and {Arbet Engels}, A. and {Arcaro}, C. and {Baack}, D. and {Babi{\'c}}, A. and {Banerjee}, B. and {Bangale}, P. and {Barres de Almeida}, U. and {Barrio}, J.~A. and {Becerra Gonz{\'a}lez}, J. and {Bednarek}, W. and {Bernardini}, E. and {Berti}, A. and {Besenrieder}, J. and {Bhattacharyya}, W. and {Bigongiari}, C. and {Biland}, A. and {Blanch}, O. and {Bonnoli}, G. and {Carosi}, R. and {Ceribella}, G. and {Cikota}, S. and {Colak}, S.~M. and {Colin}, P. and {Colombo}, E. and {Contreras}, J.~L. and {Cortina}, J. and {Covino}, S. and {D'Elia}, V. and {da Vela}, P. and {Dazzi}, F. and {de Angelis}, A. and {de Lotto}, B. and {Delfino}, M. and {Delgado}, J. and {di Pierro}, F. and {Do Souto Espi{\~n}era}, E. and {Dom{\'\i}nguez}, A. and {Dominis Prester}, D. and {Dorner}, D. and {Doro}, M. and {Einecke}, S. and {Elsaesser}, D. and {Fallah Ramazani}, V. and {Fattorini}, A. and {Fern{\'a}ndez-Barral}, A. and {Ferrara}, G. and {Fidalgo}, D. and {Foffano}, L. and {Fonseca}, M.~V. and {Font}, L. and {Fruck}, C. and {Galindo}, D. and {Gallozzi}, S. and {Garc{\'\i}a L{\'o}pez}, R.~J. and {Garczarczyk}, M. and {Gaug}, M. and {Giammaria}, P. and {Godinovi{\'c}}, N. and {Guberman}, D. and {Hadasch}, D. and {Hahn}, A. and {Hassan}, T. and {Herrera}, J. and {Hoang}, J. and {Hrupec}, D. and {Inoue}, S. and {Ishio}, K. and {Iwamura}, Y. and {Kubo}, H. and {Kushida}, J. and {Kuve{\v{z}}di{\'c}}, D. and {Lamastra}, A. and {Lelas}, D. and {Leone}, F. and {Lindfors}, E. and {Lombardi}, S. and {Longo}, F. and {L{\'o}pez}, M. and {L{\'o}pez-Oramas}, A. and {Maggio}, C. and {Majumdar}, P. and {Makariev}, M. and {Maneva}, G. and {Manganaro}, M. and {Mannheim}, K. and {Maraschi}, L. and {Mariotti}, M. and {Mart{\'\i}nez}, M. and {Masuda}, S. and {Mazin}, D. and {Minev}, M. and {Miranda}, J.~M. and {Mirzoyan}, R. and {Molina}, E. and {Moralejo}, A. and {Moreno}, V. and {Moretti}, E. and {Munar-Adrover}, P. and {Neustroev}, V. and {Niedzwiecki}, A. and {Nievas Rosillo}, M. and {Nigro}, C. and {Nilsson}, K. and {Ninci}, D. and {Nishijima}, K. and {Noda}, K. and {Nogu{\'e}s}, L. and {Paiano}, S. and {Palacio}, J. and {Paneque}, D. and {Paoletti}, R. and {Paredes}, J.~M. and {Pedaletti}, G. and {Pe{\~n}il}, P. and {Peresano}, M. and {Persic}, M. and {Prada Moroni}, P.~G. and {Prandini}, E. and {Puljak}, I. and {Garcia}, J.~R. and {Rhode}, W. and {Rib{\'o}}, M. and {Rico}, J. and {Righi}, C. and {Rugliancich}, A. and {Saha}, L. and {Saito}, T. and {Satalecka}, K. and {Schweizer}, T. and {Sitarek}, J. and {{\v{S}}nidari{\'c}}, I. and {Sobczynska}, D. and {Somero}, A. and {Stamerra}, A. and {Strzys}, M. and {Suri{\'c}}, T. and {Tavecchio}, F. and {Temnikov}, P. and {Terzi{\'c}}, T. and {Teshima}, M. and {Torres-Alb{\`a}}, N. and {Tsujimoto}, S. and {Vanzo}, G. and {Vazquez Acosta}, M. and {Vovk}, I. and {Ward}, J.~E. and {Will}, M. and {Zari{\'c}}, D. and {MAGIC Collaboration}},
        title = "{Constraining dark matter lifetime with a deep gamma-ray survey of the Perseus galaxy cluster with MAGIC}",
      journal = {Phys. Dark Univ.},
         year = 2018,
        month = dec,
       volume = {22},
        pages = {38-47},
          doi = {10.1016/j.dark.2018.08.002},
archivePrefix = {arXiv},
       eprint = {1806.11063},
 primaryClass = {astro-ph.HE},
       adsurl = {https://ui.adsabs.harvard.edu/abs/2018PDU....22...38A}
}

@article{Bergstr_m_1998,
    author = "Bergstrom, Lars and Ullio, Piero and Buckley, James H.",
    title = "{Observability of gamma-rays from dark matter neutralino annihilations in the Milky Way halo}",
    eprint = "astro-ph/9712318",
    archivePrefix = "arXiv",
    doi = "10.1016/S0927-6505(98)00015-2",
    journal = "Astropart. Phys.",
    volume = "9",
    pages = "137--162",
    year = "1998"
}

@article{Frumkin_2023,
    author = "Frumkin, Ronny and Kuflik, Eric and Lavie, Itay and Silverwater, Tal",
    title = "{Roadmap to Thermal Dark Matter beyond the Weakly Interacting Dark Matter Unitarity Bound}",
    eprint = "2207.01635",
    archivePrefix = "arXiv",
    primaryClass = "hep-ph",
    doi = "10.1103/PhysRevLett.130.171001",
    journal = "Phys. Rev. Lett.",
    volume = "130",
    number = "17",
    pages = "171001",
    year = "2023"
}

@ARTICLE{Navarro_1996,
       author = {{Navarro}, Julio F. and {Frenk}, Carlos S. and {White}, Simon D.~M.},
        title = "{The Structure of Cold Dark Matter Halos}",
      journal = {Astrophys.\ J.},
         year = 1996,
        month = may,
       volume = {462},
        pages = {563},
          doi = {10.1086/177173},
archivePrefix = {arXiv},
       eprint = {astro-ph/9508025},
 primaryClass = {astro-ph},
       adsurl = {https://ui.adsabs.harvard.edu/abs/1996ApJ...462..563N}
}

@ARTICLE{Retana_Montenegro_2012,
    author = "Retana-Montenegro, E. and Van Hese, E. and Gentile, G. and Baes, M. and Frutos-Alfaro, F.",
    title = "{Analytical properties of Einasto dark matter haloes}",
    eprint = "1202.5242",
    archivePrefix = "arXiv",
    primaryClass = "astro-ph.CO",
    doi = "10.1051/0004-6361/201118543",
    journal = "Astron. Astrophys.",
    volume = "540",
    pages = "A70",
    year = "2012"
}

@ARTICLE{Einasto_1965,
       author = {{Einasto}, J.},
        title = "{On the Construction of a Composite Model for the Galaxy and on the Determination of the System of Galactic Parameters}",
      journal = {Trudy Astrofizicheskogo Instituta Alma-Ata},
         year = 1965,
        month = jan,
       volume = {5},
        pages = {87-100},
       adsurl = {https://ui.adsabs.harvard.edu/abs/1965TrAlm...5...87E}
}

@ARTICLE{Hernquist_1990,
    author = "Hernquist, Lars",
    title = "{An Analytical Model for Spherical Galaxies and Bulges}",
    reportNumber = "IASSNS-AST-89-63",
    doi = "10.1086/168845",
    journal = "Astrophys. J.",
    volume = "356",
    pages = "359",
    year = "1990"
}

@article{DiCintio_2013,
    author = "Di Cintio, Arianna and Brook, Chris B. and Macci{\`o}, Andrea V. and Stinson, Greg S. and Knebe, Alexander and Dutton, Aaron A. and Wadsley, James",
    title = "{The dependence of dark matter profiles on the stellar-to-halo mass ratio: a prediction for cusps versus cores}",
    eprint = "1306.0898",
    archivePrefix = "arXiv",
    primaryClass = "astro-ph.CO",
    doi = "10.1093/mnras/stt1891",
    journal = "Mon. Not. Roy. Astron. Soc.",
    volume = "437",
    number = "1",
    pages = "415--423",
    year = "2014"
}

@article{Errani_2021,
    author = {Errani, Rapha{\"e}l and Navarro, Julio F.},
    title = "{The asymptotic tidal remnants of cold dark matter subhaloes}",
    eprint = "2011.07077",
    archivePrefix = "arXiv",
    primaryClass = "astro-ph.GA",
    doi = "10.1093/mnras/stab1215",
    journal = "Mon. Not. Roy. Astron. Soc.",
    volume = "505",
    number = "1",
    pages = "18--32",
    year = "2021"
}

@ARTICLE{Marco_Cirelli_2011,
       author = {{Cirelli}, Marco and {Corcella}, Gennaro and {Hektor}, Andi and {H{\"u}tsi}, Gert and {Kadastik}, Mario and {Panci}, Paolo and {Raidal}, Martti and {Sala}, Filippo and {Strumia}, Alessandro},
        title = "{PPPC 4 DM ID: a poor particle physicist cookbook for dark matter indirect detection}",
      journal = {JCAP},
         year = 2011,
        month = mar,
       volume = {2011},
       number = {3},
          eid = {051},
        pages = {051},
          doi = {10.1088/1475-7516/2011/03/051},
archivePrefix = {arXiv},
       eprint = {1012.4515},
 primaryClass = {hep-ph},
       adsurl = {https://ui.adsabs.harvard.edu/abs/2011JCAP...03..051C}
}

@ARTICLE{Ciafaloni_2011,
       author = {{Ciafaloni}, Paolo and {Comelli}, Denis and {Riotto}, Antonio and {Sala}, Filippo and {Strumia}, Alessandro and {Urbano}, Alfredo},
        title = "{Weak corrections are relevant for dark matter indirect detection}",
      journal = {JCAP},
         year = 2011,
        month = mar,
       volume = {03},
       number = {3},
          eid = {019},
        pages = {019},
          doi = {10.1088/1475-7516/2011/03/019},
archivePrefix = {arXiv},
       eprint = {1009.0224},
 primaryClass = {hep-ph},
       adsurl = {https://ui.adsabs.harvard.edu/abs/2011JCAP...03..019C}
}

@ARTICLE{Fischer_2016,
       author = {{Fischer}, N. and {Prestel}, S. and {Ritzmann}, M. and {Skands}, P.},
        title = "{VINCIA for hadron colliders}",
      journal = {Eur. Phys. J. C},
         year = 2016,
        month = nov,
       volume = {76},
       number = {11},
          eid = {589},
        pages = {589},
          doi = {10.1140/epjc/s10052-016-4429-6},
archivePrefix = {arXiv},
       eprint = {1605.06142},
 primaryClass = {hep-ph},
       adsurl = {https://ui.adsabs.harvard.edu/abs/2016EPJC...76..589F}
}

@ARTICLE{Amoroso_2019,
       author = {{Amoroso}, Simone and {Caron}, Sascha and {Jueid}, Adil and {Ruiz de Austri}, Roberto and {Skands}, Peter},
        title = "{Estimating QCD uncertainties in Monte Carlo event generators for gamma-ray dark matter searches}",
      journal = {JCAP},
         year = 2019,
        month = may,
       volume = {2019},
       number = {5},
          eid = {007},
        pages = {007},
          doi = {10.1088/1475-7516/2019/05/007},
archivePrefix = {arXiv},
       eprint = {1812.07424},
 primaryClass = {hep-ph},
       adsurl = {https://ui.adsabs.harvard.edu/abs/2019JCAP...05..007A}
}

@ARTICLE{breit1934a,
       author = {{Breit}, G. and {Wheeler}, John A.},
        title = "{Collision of Two Light Quanta}",
      journal = {Phys. Rev.},
         year = 1934,
        month = dec,
       volume = {46},
       number = {12},
        pages = {1087-1091},
          doi = {10.1103/PhysRev.46.1087},
       adsurl = {https://ui.adsabs.harvard.edu/abs/1934PhRv...46.1087B}
}

@ARTICLE{klein1929,
       author = {{Klein}, O. and {Nishina}, T.},
        title = "{{\"U}ber die Streuung von Strahlung durch freie Elektronen nach der neuen relativistischen Quantendynamik von Dirac}",
      journal = {Zeitschrift fur Physik},
         year = 1929,
        month = nov,
       volume = {52},
       number = {11-12},
        pages = {853-868},
          doi = {10.1007/BF01366453},
       adsurl = {https://ui.adsabs.harvard.edu/abs/1929ZPhy...52..853K}
}

@article{Cheng1970,
    author = "Cheng, Hung and Wu, Tai Tsun",
    title = "{Cross sections for two-pair production at infinite energy}",
    doi = "10.1103/PhysRevD.2.2103",
    journal = "Phys. Rev. D",
    volume = "2",
    pages = "2103--2105",
    year = "1970"
}

@article{Bonometto1972,
    author = "Bonometto, S. A. and Marcolungo, P.",
    title = "{Metagalactic opacity to photons of energy larger than 10-to-the-17 ev}",
    doi = "10.1007/BF02752747",
    journal = "Lett. Nuovo Cim.",
    volume = "5S2",
    pages = "595--603",
    year = "1972"
}

@ARTICLE{Heiter_2018,
       author = {{Heiter}, C. and {Kuempel}, D. and {Walz}, D. and {Erdmann}, M.},
        title = "{Production and propagation of ultra-high energy photons using CRPropa 3}",
      journal = {Astropart. Phys.},
         year = 2018,
        month = nov,
       volume = {102},
        pages = {39-50},
          doi = {10.1016/j.astropartphys.2018.05.003},
archivePrefix = {arXiv},
       eprint = {1710.11406},
 primaryClass = {astro-ph.IM},
       adsurl = {https://ui.adsabs.harvard.edu/abs/2018APh...102...39H}
}

@ARTICLE{Alves_Batista_2021,
       author = {{Alves Batista}, Rafael and {Saveliev}, Andrey},
        title = "{The Gamma-Ray Window to Intergalactic Magnetism}",
      journal = {Universe},
         year = 2021,
        month = jul,
       volume = {7},
       number = {7},
          eid = {223},
        pages = {223},
          doi = {10.3390/universe7070223},
archivePrefix = {arXiv},
       eprint = {2105.12020},
 primaryClass = {astro-ph.HE},
       adsurl = {https://ui.adsabs.harvard.edu/abs/2021Univ....7..223A}
}

@ARTICLE{Saldana_Lopez_2021,
       author = {{Saldana-Lopez}, Alberto and {Dom{\'\i}nguez}, Alberto and {P{\'e}rez-Gonz{\'a}lez}, Pablo G. and {Finke}, Justin and {Ajello}, Marco and {Primack}, Joel R. and {Paliya}, Vaidehi S. and {Desai}, Abhishek},
        title = "{An observational determination of the evolving extragalactic background light from the multiwavelength HST/CANDELS survey in the Fermi and CTA era}",
      journal = {Mon. Not. Roy. Astron. Soc.},
         year = 2021,
        month = nov,
       volume = {507},
       number = {4},
        pages = {5144-5160},
          doi = {10.1093/mnras/stab2393},
archivePrefix = {arXiv},
       eprint = {2012.03035},
 primaryClass = {astro-ph.CO},
       adsurl = {https://ui.adsabs.harvard.edu/abs/2021MNRAS.507.5144S}
}

@ARTICLE{NITU2021102532,
       author = {{Ni{\c{t}}u}, I.~C. and {Bevins}, H.~T.~J. and {Bray}, J.~D. and {Scaife}, A.~M.~M.},
        title = "{An updated estimate of the cosmic radio background and implications for ultra-high-energy photon propagation}",
      journal = {Astropart. Phys.},
         year = 2021,
        month = mar,
       volume = {126},
          eid = {102532},
        pages = {102532},
          doi = {10.1016/j.astropartphys.2020.102532},
archivePrefix = {arXiv},
       eprint = {2004.13596},
 primaryClass = {astro-ph.HE},
       adsurl = {https://ui.adsabs.harvard.edu/abs/2021APh...12602532N}
}

@ARTICLE{kang2024_perseus,
       author = {{Kang}, Wooseok and {Hwang}, Ho Seong and {Song}, Hyunmi and {Park}, Changbom and {Hwang}, Narae and {Park}, Byeong-Gon},
        title = "{A Deep Redshift Survey of the Perseus Cluster (A426): Spatial Distribution and Kinematics of Galaxies}",
      journal = {Astrophys.\ J. Suppl.},
         year = 2024,
        month = may,
       volume = {272},
       number = {1},
          eid = {22},
        pages = {22},
          doi = {10.3847/1538-4365/ad390d},
archivePrefix = {arXiv},
       eprint = {2403.19307},
 primaryClass = {astro-ph.GA},
       adsurl = {https://ui.adsabs.harvard.edu/abs/2024ApJS..272...22K}
}

@article{Blumenthal1970,
  title = {Bremsstrahlung, Synchrotron Radiation, and Compton Scattering of High-Energy Electrons Traversing Dilute Gases},
  author = {BLUMENTHAL, GEORGE R. and GOULD, ROBERT J.},
  journal = {Rev. Mod. Phys.},
  volume = {42},
  issue = {2},
  pages = {237--270},
  numpages = {0},
  year = {1970},
  month = {Apr},
  publisher = {American Physical Society},
  doi = {10.1103/RevModPhys.42.237},
  url = {https://link.aps.org/doi/10.1103/RevModPhys.42.237}
}

\end{document}